\documentclass[aps,prd,superscriptaddress,nofootinbib]{revtex4-2} 
\usepackage{amsmath,amssymb,booktabs,graphicx,microtype}

\usepackage[caption=false]{subfig}
\usepackage{placeins}
\usepackage[hidelinks,pdftitle={From a Sharp Thin-Shell Obstruction to a Smooth Positive-Density Initial-Data Embedding in Lambda-FLRW Cosmology},pdfauthor={Seokcheon Lee}]{hyperref}
\usepackage{orcidlink}
\usepackage{xcolor}

\newcommand{\dd}{\mathrm d}
\newcommand{\Rvir}{R_{\rm vir}}
\newcommand{\Rcomp}{R_{\rm comp}}
\newcommand{\ee}{\varepsilon}

\begin{document}

\title{From a Sharp Thin-Shell Obstruction to a Smooth Positive-Density Initial-Data Embedding of a Virialized Halo in \texorpdfstring{$\boldsymbol{\Lambda}$}{Lambda}--FLRW Cosmology}

\author{Seokcheon Lee \orcidlink{0000-0003-0861-1300}}
\email{skylee@skku.edu}
\affiliation{Department of Physics, Institute of Basic Science, Sungkyunkwan University, Suwon 16419, Korea}

\begin{abstract}
Within classical general relativity, we compare a sharp timelike junction and a smooth finite-width spacelike initial-data embedding of a positive-excess virialized halo in a homogeneous $\Lambda$–FLRW environment. On the ordinary branch, timelike Israel matching at the virial boundary produces a negative surface layer: a distributional representation of the environmental compensation omitted by the sharp construction. We replace this zero-width source by a finite-width, background-relative underdensity whose local rest-frame energy density remains positive. The resulting conformally flat, constant-mean-curvature ADM initial slice satisfies the Hamiltonian and momentum constraints, with the residuals converging under resolution refinement. At finite radius, its geometric and matter variables return to the local-effective FLRW data within the declared tolerances; the constructed slice contains no residual distributional layer and satisfies the reconstructed weak, null, and dominant energy conditions. For radii beyond the numerical endpoint, the exterior is defined by the exact analytic local-effective FLRW solution. The GCT framework supplies the global motivation and clock interpretation, while the present calculations are local-effective classical-GR calculations with the bound-sector constants $c_0$ and $G_0$ held fixed; no radial GCT lapse or constant interpolation is constructed.
\end{abstract}

\maketitle

\tableofcontents

\section{Introduction}
\label{sec:introduction}

In classical general relativity (GR), the embedding question studied here can be addressed through two geometrically distinct formulations based on standard methods. A sharp construction joins two spacetime regions across a timelike world tube and applies the Darmois--Israel junction conditions~\cite{Israel:1966rt}. In cosmological Swiss-cheese constructions, smooth matching without a surface layer is known to impose restrictive boundary and mass-matching conditions~\cite{Einstein:1945id,Carrera:2008pi,Grenon:2011fs}. A smooth construction instead uses the conformal initial-data method~\cite{York:1978gql,Maxwell:2014qpa} to solve the Hamiltonian and momentum constraints on one spherical conformal ADM slice, and then checks finite-radius FLRW closure, local energy conditions, and the Misner--Sharp invariant. For the specific local-static--to--local-effective-FLRW matching considered here, the sharp construction requires a negative surface layer on the ordinary branch; the smooth construction asks whether finite-width positive-density initial data can replace that distributional source. Neither construction, by itself, is a time evolution.

Gravitationally bound systems do not generally follow the background Hubble expansion in their internal length scales: local rulers are not obtained by simply multiplying bound-system lengths by the cosmological scale factor~\cite{Carrera:2008pi}. The Robertson--Walker (RW) construction rests on the cosmological principle (CP) together with Weyl's postulate. Spatial homogeneity and isotropy restrict the constant-time hypersurfaces to maximally symmetric three-spaces, while Weyl's postulate identifies a nonintersecting timelike congruence of fundamental observers whose worldlines define the comoving cosmic medium and are orthogonal to those hypersurfaces~\cite{Katanaev:2016byi,Islam:1992nt,Hobson:2006se}. Before fixing the parametrization along this congruence, the metric may be written as
\[
 ds^2 = -N_{\rm bg}^2(T)c_0^2\,dT^2 + a^2(T)\gamma_{ij}dx^i dx^j .
\]
The spatial Killing symmetries determine the form of $\gamma_{ij}$, and Weyl's postulate identifies the preferred comoving congruence, but neither condition selects a unique parametrization of time along that congruence. The conventional choice $N_{\rm bg}=1$ is therefore a valid synchronous cosmological proper-time gauge, rather than a condition uniquely imposed by the CP or by Weyl's postulate.

At the level of the homogeneous metric alone, a time-dependent lapse can be removed by a reparametrization of the cosmological time coordinate. The
generalized cosmological time (GCT) framework asks the additional operational question of whether the resulting global cosmological clock normalization must coincide with the proper-time normalization of every gravitationally bound system and with that sampled by cosmological propagation and timing observables~\cite{Lee:2025osx,Lee:2026sow}. In GCT, the homogeneous cosmological sector may carry an evolving lapse, whereas a locally bound sector is described in its own proper time with
\[
 N_{\rm loc}=1, \qquad c_{\rm loc}=c_0, \qquad G_{\rm loc}=G_0.
\]
The corresponding background scalings and density conventions are collected in Sec.~\ref{sec:gct-overview} rather than repeated here.

This distinction has observational motivation. Results from Type Ia supernovae, gamma-ray bursts (GRBs), and quasars need not imply different background time-dilation laws for different source classes~\cite{Lee:2023ucu,Lee:2024kxa,Lee:2026kyz}. Supernova and GRB transient durations can track the background propagation factor relatively directly, whereas fixed-observed-band quasar variability can also include redshift-dependent source-region selection and intrinsic time shortening. This phenomenology motivates separating global cosmological timing from local bound clocks. The present paper does not revisit those data. It addresses the prior geometric question of whether the assumed local clock sector can be embedded in a cosmological environment without an unphysical singular layer.

Earlier GCT studies motivated gravitationally bound and virialized systems as local proper-time sectors in the interpretation of cosmological timing observables~\cite{Lee:2026kyz}. Our subsequent work addressed the coexistence of the global and local sectors as an idealized geometric junction problem~\cite{Lee:2026sow}. That analysis represented the local--cosmological separation by a zero-width timelike matching and established a geometric consistency condition for embedding a locally static region in the cosmological background.

The present work asks the next physical question: whether that zero-width idealization can be replaced by a physically admissible finite-width environment associated with a virialized bound system. A realistic virialized halo is generally surrounded by an extended transition and infall environment rather than immediately by an exactly homogeneous cosmological medium. In the finite-radius embedding considered here, this environment also supplies the compensating deficit required for closure. Therefore, we test whether such an environment can replace the singular source generated by sharp matching. We show constructively that a positive-density, constraint-satisfying relativistic initial-data embedding exists, connecting the assumed virialized core through a finite compensating region to a homogeneous local-effective FLRW exterior at finite radius.

This route is conceptually distinct from Vainshtein- or chameleon-type screening~\cite{Vainshtein:1972sx,Khoury:2003aq,Khoury:2003rn}. The Vainshtein mechanism suppresses an additional gravitational degree of freedom through nonlinear derivative interactions inside a source-dependent radius, whereas the chameleon mechanism relies on a scalar field whose effective mass depends on the ambient matter density. Neither type of structure is introduced here: the construction adds no new field or propagating gravitational degree of freedom, no nonlinear derivative screening radius, and no environment-dependent scalar effective mass. 

The present construction is therefore not a modified-gravity screening model. It is formulated as a local-effective classical-GR calculation with the bound-sector constants $c_0$ and $G_0$ held fixed. Gravitational collapse, binding, and virialization instead provide a standard structure-formation criterion for identifying an assumed bound region~\cite{Lahav:1991wc,Lee:2009dz,Lee:2009qq}. In the GCT interpretation adopted here, virialization is used only as a working environmental criterion for associating that region with the assumed local proper-time sector, without introducing an additional scalar field or screening prescription. The present work tests only the initial-data admissibility of that sector as a prerequisite for any subsequent dynamical analysis. Its formation, persistence, and stability require a separate numerical evolution treatment and are not established here.

A virialized halo carries positive excess mass relative to its background. If that excess is cut off at the virial radius and matched directly to an exactly homogeneous exterior, the compensating deficit required for finite-radius mass closure is omitted. For the ordinary sharp matching considered here, the omitted compensation is represented by a negative distributional source on the matching surface. The central question is whether a positive-excess virialized core can instead be connected, on a single spacelike initial slice, to a homogeneous local-effective Friedmann--Lema{\^\i}tre--Robertson--Walker (FLRW) exterior through a finite-width, everywhere positive, constraint-satisfying compensating environment.

We first identify the obstruction produced by sharp ordinary matching and relate it to the mass-locking logic of the Einstein--Straus vacuole, the Darmois--Israel junction conditions, and related cosmological thin-shell constructions~\cite{Einstein:1945id,Israel:1966rt,Grenon:2011fs,Sahu:2024flg}. Then, we construct a smooth, conformally flat, constant-mean-curvature initial slice containing an assumed virialized positive-excess core and a finite compensating environment. This provides the finite-width initial-data bridge absent from the earlier idealized junction construction.

This work contains two geometrically distinct calculations. The sharp calculation joins the local-static boundary representation to the local-effective FLRW exterior across a timelike world tube and diagnoses the resulting Israel layer. The smooth calculation does not replace that world tube by another timelike junction. It instead constructs one spacelike conformal ADM initial slice extending from the assumed virialized core through a finite compensating region to the local-effective FLRW exterior. The global GCT background provides the cosmological interpretation of these local-effective calculations, but the present work constructs neither a full timelike junction to that background nor a radial interpolation of the lapse, $c$, and $G$.

The paper is organized as follows. Sec.~\ref{sec:gct-overview} fixes the GCT
and local-effective bookkeeping, Sec.~\ref{sec:virial} supplies the virial
boundary data, and Sec.~\ref{sec:sharp-matching} derives and interprets the
sharp obstruction. Section~\ref{sec:smooth} presents the smooth initial-data
construction, positive solution, finite-radius closure, and mass and matter
diagnostics. Section~\ref{sec:validation} collects the independent numerical
validation, physical interpretation, and precise claim boundaries, followed
by Sec.~\ref{sec:conclusions}. All physical equations use SI units; the
dimensionless variables and restoration map are given in Appendix~\ref{app:conventions-normalization}.

\section{Geometric framework and the two constructions}
\label{sec:gct-overview} 

This section distinguishes the geometric descriptions used in the paper; they are not separate dynamical phases derived from one another. Space I denotes the full homogeneous GCT background, written with a nontrivial cosmological lapse. At the homogeneous level, Space II-A is the same expanding exterior expressed in a proper-time, local-effective representation with the fixed reference constants $c_0$ and $G_0$. It is not a separate $b=0$ cosmology, nor does it introduce an independent local value of the GCT parameter $b$.

Space II-B is the numerically constructed conformal ADM portion extending from the assumed virialized-core data through the finite compensation region to $R_{\rm comp}$, where it closes to the Space II-A FLRW data within the declared tolerances. Beyond $R_{\rm comp}$, the same spacelike hypersurface is continued analytically by the exact Space II-A exterior. Space III denotes the locally bound sector: Space III-A specifies the local proper-time and assumed virialized-core interpretation, whereas Space III-B is the local-static exterior boundary representation used only in the sharp junction calculation. Thus, in the smooth construction, the complete initial hypersurface consists of the numerical Space II-B portion and its exact analytic Space II-A continuation beyond $R_{\rm comp}$.

\begin{figure}[!htbp]
\centering
\includegraphics[width=0.90\linewidth]{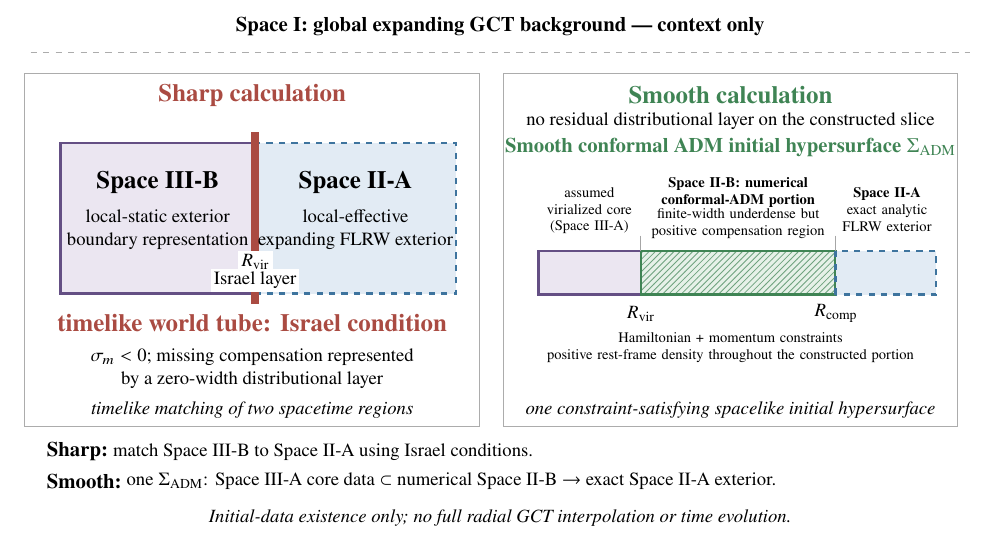}
\caption{Geometric distinction between the sharp and smooth calculations. The sharp calculation joins the local-static exterior boundary representation of Space III-B to the local-effective FLRW exterior of Space II-A across a timelike world tube; on the ordinary branch, omitted finite environmental compensation is represented by a negative Israel layer. The smooth calculation constructs the numerical Space II-B portion of a conformal ADM initial hypersurface. Its inner data represent the assumed Space III-A virialized-core sector, and its finite-width compensation region remains everywhere positive. At $R_{\rm comp}$, the numerical Space II-B data close to Space II-A within the declared tolerances; beyond that radius, the hypersurface is continued by the exact analytic Space II-A exterior. Space I supplies the global GCT context, while Space II-A is its homogeneous proper-time, local-effective representation used in the calculations. No radial profile connecting the global and locally bound clock descriptions is constructed here.}
\label{fig:conceptualarchitecture}
\end{figure}

Figure~\ref{fig:conceptualarchitecture} summarizes this hierarchy and the geometric distinction between the sharp timelike junction and the smooth spacelike initial-data construction.

\subsection{Space I: global homogeneous GCT background}
\label{sec:gct-homogeneous} 

The homogeneous GCT background is written as
\begin{equation}
 ds^2 = -N_{\rm bg}^2(T)c_0^2\,dT^2 + a^2(T)\gamma_{ij}dx^i dx^j , \label{eq:globalmetric}
\end{equation}
with the meVSL lapse parametrization
\begin{equation}
 N_{\rm bg}(a)=a^{b/4}. \label{eq:gct-lapse-ansatz}
\end{equation}
Here $T$ is the adopted global cosmological time coordinate and $b$ parametrizes the corresponding homogeneous clock rate. The parameter $b$ is not an independent propagating field. In the meVSL/GCT formulation, variation with respect to the associated time-dependent clock coefficient does not yield a dynamical equation for that coefficient or for $b$; instead, it imposes a constraint on the background geometry~\cite{Lee:2025osx}. Consequently, the value of $b$ specifies the adopted homogeneous lapse parametrization and may be constrained by cosmological observations.

It is convenient to define the background clock coefficient
\begin{equation}
 c_{\rm bg}(a) \equiv N_{\rm bg}(a)c_0 = c_0a^{b/4}. \label{eq:cbg-definition}
\end{equation}
This quantity is the coefficient relating the chosen global time coordinate to the homogeneous proper-time interval; it is not an independently propagating local light-speed field.

Within the correlated meVSL scaling prescription, the lapse assignment is accompanied by
\begin{equation}
 c_{\rm bg}=c_0a^{b/4}, \qquad G_{\rm bg}=G_0a^b, \qquad m_{\rm bg}=m_0a^{-b/2}, \qquad \hbar_{\rm bg}=\hbar_0a^{-b/4}.
\label{eq:gct-background-scalings}
\end{equation}
We adopt $a_0=1$, so that the subscript $0$ denotes the present-epoch reference values of the homogeneous background:
\begin{equation}
 c_{\rm bg}(a_0)=c_0, \qquad G_{\rm bg}(a_0)=G_0, \qquad m_{\rm bg}(a_0)=m_0, \qquad \hbar_{\rm bg}(a_0)=\hbar_0.
\label{eq:present-reference-values}
\end{equation}
Within the GCT interpretation adopted below, the same $c_0$ and $G_0$ also serve as the fixed reference constants used to describe the assumed locally bound proper-time sector.

The corresponding electromagnetic and thermodynamic scalings are described in Refs.~\cite{Lee:2020zts,Lee:2023bjz,Lee:2022heb,Lee:2024zcu}. The correlated assignments preserve the standard local forms of special relativity, electromagnetism, quantum mechanics, and thermodynamics on each homogeneous hypersurface. In particular,
\begin{equation}
 m_{\rm bg}c_{\rm bg}^2=m_0c_0^2, \qquad \kappa = \frac{8\pi G_{\rm bg}}{c_{\rm bg}^4} = \frac{8\pi G_0}{c_0^4}. \label{eq:gct-kappa-invariant}
\end{equation}

These relations describe correlated homogeneous-background quantities. They are not interpreted as independently drifting constants measured by an observer in the assumed locally bound sector. In the local-effective calculations below, $c_0$, $G_0$, $\hbar$ and the associated atomic standards are held fixed. Since the lapse parametrization introduces no additional propagating scalar degree of freedom, the present construction does not invoke a Vainshtein-, chameleon-, or analogous field-screening mechanism.

Space I supplies the global cosmological context for both calculations. The sharp and smooth constructions, however, use the homogeneous proper-time Space II-A representation introduced below, rather than the full global-coordinate description. Constructing the inhomogeneous radial and temporal geometry that connects the cosmological clock to a locally bound proper-time sector---including any profiles $N(T,r)$, $c(T,r)$, or $G(T,r)$---and testing its formation and persistence require a separate numerical evolution.

\subsection{Space II-A: local-effective FLRW exterior}
\label{sec:gct-local-effective} 

The exterior reference used by both calculations is an expanding, spatially flat local-effective FLRW geometry. Space II-A is obtained by expressing the homogeneous Space I exterior in the proper-time coordinate $T_+$, defined by
\begin{equation}
 dT_+ = N_{\rm bg}(T)\,dT = \frac{c_{\rm bg}(T)}{c_0}\,dT \quad \Rightarrow \quad c_0 dT_+ = c_{\rm bg}(T) dT .\label{eq:proper-time-map}
\end{equation}
Equation~\eqref{eq:proper-time-map} makes explicit that the two descriptions assign the same homogeneous proper-time interval, $c_0 dT_+ = c_{\rm bg} dT$. Dividing this relation by $a$ gives 
\begin{equation}
 \frac{c_{\rm bg}\,dT}{a} = \frac{c_0\,dT_+}{a}.
\end{equation}
Hence the radial null-propagation element is unchanged by passing from the global-time description to the proper-time representation. Thus, Space II-A is not obtained by changing the homogeneous causal geometry, but by expressing it with the proper-time clock normalization. Accordingly, the homogeneous metric becomes
\begin{equation}
 ds_+^2 = -c_0^2\,dT_+^2 +a^2(T_+) \left( dr^2+r^2d\Omega^2 \right). \label{eq:local-effective-flrw-metric}
\end{equation}

The label local-effective (LE) denotes the homogeneous proper-time representation used in the local calculations, with $c_0$ and $G_0$ as the fixed reference constants. Algebraically, its density relations take the standard forms obtained when the explicit $b$-dependent factors are removed. This does not define a separate parameter $b_{\rm LE}$, a new cosmological solution, or a dynamical transition of $b$.

Define the expansion rates with respect to the two homogeneous time coordinates by
\begin{equation}
 H_T \equiv  \frac{1}{a}\frac{da}{dT},  \qquad H_p \equiv \frac{1}{a}\frac{da}{dT_+}. \label{eq:hubble-coordinate-definitions}
\end{equation}
Equation~\eqref{eq:proper-time-map} gives
\begin{equation}
 H_p = \frac{H_T}{N_{\rm bg}} = \frac{c_0}{c_{\rm bg}}H_T \quad \Rightarrow \quad \frac{c_0}{H_p} = \frac{c_{\rm bg}}{H_T} . \label{eq:hubble-coordinate-map}
\end{equation}
Thus, the Hubble length is representation independent. The lapse changes the numerical values assigned separately to the clock coefficient and the Hubble rate, but not their ratio. This is the same cancellation emphasized in the homogeneous meVSL formulation, where the Hubble radius agrees with that of the corresponding standard proper-time FLRW description~\cite{Lee:2025osx}. In this respect the lapse-based meVSL/GCT construction differs from early-time VSL scenarios that invoke a genuinely enlarged $c/H$ or causal horizon: the lapse reparametrization itself does not enlarge the homogeneous Hubble length.

Introducing the areal radius,
\begin{equation}
 R=a(T_+)r, \qquad dR=H_pR\,dT_+ +a\,dr , \label{eq:areal-radius-transform}
\end{equation}
the same metric given in Eq.~\eqref{eq:local-effective-flrw-metric} takes the form
\begin{equation}
 ds_+^2  = -c_0^2\,dT_+^2 +\left(dR-H_pR\,dT_+\right)^2 +R^2d\Omega^2 . \label{eq:flrw-areal-radius-metric}
\end{equation}
The $H_pR$ cross term remains: Space II-A is expanding rather than static.

The invariant norm of the areal-radius gradient is
\begin{align}
 F_+ \equiv g_+^{ab}\partial_aR\,\partial_bR = 1-\frac{H_p^2R^2}{c_0^2} = 1-\frac{H_T^2R^2}{N_{\rm bg}^2c_0^2}
 = 1-\frac{H_T^2R^2}{c_{\rm bg}^2}. \label{eq:fpluslocal}
\end{align}
Thus, the proper-time and global-coordinate expressions describe the same homogeneous invariant when the transformation of the expansion rate is included consistently. In the verified sharp calculation and at the exact analytic exterior of the smooth construction, this invariant is evaluated in the Space II-A proper-time form,
\begin{equation}
 F_+^{\rm eff} = 1-\frac{H_p^2R^2}{c_0^2}. \label{eq:fplus-effective-form}
\end{equation}
The distinction between Spaces I and II-A is therefore not a difference in the homogeneous areal-radius invariant, but the representation used in the local calculation and the absence, in the present work, of an inhomogeneous radial interpolation to the bound sector.

The density bookkeeping follows the same distinction. Here the subscript $m$ denotes pressureless matter; radiation is not included in the low-redshift collapse and compensation benchmark used below. In the homogeneous Space I description, the background mass density of pressureless matter and its corresponding energy density are
\begin{align}
 \rho_{m,\rm bg}(a) &= \rho_{m0}a^{-3-b/2}, & e_{m,b}(a) &\equiv \rho_{m,\rm bg}(a)c_{\rm bg}^2(a) = \rho_{m0}c_0^2a^{-3},
\label{eq:matterdensitymap}
\end{align}
as derived in Refs.~\cite{Lee:2020zts,Lee:2025osx}. The explicit $b$-dependence cancels in the energy density because the mass-density and clock-coefficient scalings are correlated.

The corresponding local-equivalent mass-density representation is
\begin{equation}
 \rho_{m,\rm LE}(a) \equiv \frac{e_{m,b}(a)}{c_0^2} = \rho_{m0}a^{-3}. \label{eq:matterLE}
\end{equation}
The LE label denotes the representation of the same homogeneous matter energy using the fixed reference conversion $c_0^2$. It does not introduce a new matter component, an independent continuity law, or a second value of the parameter $b$.

The corresponding vacuum quantities are
\begin{align}
 \rho_{\Lambda,\rm bg}(a) &= \frac{\Lambda c_{\rm bg}^2(a)}{8\pi G_{\rm bg}(a)} = \rho_{\Lambda0}a^{-b/2}, \nonumber\\
 e_{\Lambda,b} &\equiv \rho_{\Lambda,\rm bg}(a)c_{\rm bg}^2(a) = \frac{\Lambda c_{\rm bg}^4(a)}{8\pi G_{\rm bg}(a)} =
 \frac{\Lambda c_0^4}{8\pi G_0} \equiv e_\Lambda, \nonumber\\
 \rho_{\Lambda,\rm LE} &\equiv \frac{e_\Lambda}{c_0^2} = \rho_{\Lambda0}, \qquad
  \rho_{\Lambda,\rm loc} \equiv \rho_{\Lambda,\rm LE}. \label{eq:vacuumdensitymap}
\end{align}

For a spatially flat homogeneous reference background containing pressureless matter and $\Lambda$, the global-coordinate Friedmann equation can be written as
\begin{align}
 H_T^2 = \frac{8\pi G_{\rm bg}}{3}\rho_{m,\rm bg} +\frac{\Lambda c_{\rm bg}^2}{3} =
 N_{\rm bg}^2 \left( \frac{8\pi G_0}{3}\rho_{m,\rm LE} +\frac{\Lambda c_0^2}{3}  \right).
\label{eq:friedmann-global}
\end{align}
Using Eq.~\eqref{eq:hubble-coordinate-map}, the equivalent Space II-A proper-time form is
\begin{equation}
 H_p^2 = \frac{8\pi G_0}{3}\rho_{m,\rm LE} +\frac{\Lambda c_0^2}{3} = \frac{8\pi G_0}{3c_0^2}e_{m,b} +\frac{\Lambda c_0^2}{3}. \label{eq:friedmann}
\end{equation}
Equations~\eqref{eq:friedmann-global} and \eqref{eq:friedmann} describe the same homogeneous expansion using different time coordinates and correlated density representations. Equation~\eqref{eq:friedmann} is the reference expansion equation used in the local-effective construction; it is not the nonlinear evolution equation of the virializing core.

The spherical-collapse and virial relations that supply the assumed core boundary data are introduced separately in Sec.~\ref{sec:virial}, following Refs.~\cite{Lahav:1991wc,Lee:2009dz,Lee:2009qq}. For later use in those relations, define
\begin{equation}
 Q(a) \equiv \frac{\rho_{m,\rm bg}}{\rho_{\Lambda,\rm bg}} = \frac{\rho_{m,\rm LE}}{\rho_{\Lambda,\rm LE}} = \frac{\rho_{m0}}{\rho_{\Lambda0}}a^{-3},
 \qquad Q_{\rm ta}\equiv Q(a_{\rm ta}), \label{eq:Qdefinition}
\end{equation}
and the turnaround overdensity
\begin{equation}
 \zeta \equiv \left.  \frac{\rho_{\rm cl,bg}}{\rho_{m,\rm bg}} \right|_{\rm ta}. \label{eq:zetadefinition}
\end{equation}
Because the cluster and background matter densities acquire the same positive conversion factor, $\zeta$, and hence $\zeta Q_{\rm ta}$, are representation independent. Compensation integrals in the local-effective construction use $\rho_{m,\rm LE}$, or the Eulerian energy density divided by $c_0^2$, rather than inserting $\rho_{m,\rm bg}$ directly.

\subsection{Space II-B: smooth conformal ADM initial slice}
\label{sec:smooth-space-roadmap} 

Space II-A denotes the four-dimensional homogeneous local-effective FLRW exterior introduced above. By contrast, Space II-B does not denote a second four-dimensional spacetime solution: it denotes the numerically constructed portion of a single spacelike initial hypersurface.

With the length-like ADM time coordinate normalized by $x^0=c_0T_+$ in the homogeneous Space II-A exterior, the general ADM context is
\begin{equation}
 \dd s^2=-\alpha^2(\dd x^0)^2 +\gamma_{ij}(\dd x^i+\beta^i\dd x^0)(\dd x^j+\beta^j\dd x^0). \label{eq:3p1metric}
\end{equation}
Here the smooth construction selects a single spacelike initial hypersurface,
\begin{equation}
 x^0=x^0_\ast,
\end{equation}
on which the radial coordinate $r$ varies. The spatial metric is taken to be conformally flat,
\begin{equation}
 \gamma_{ij}\dd x^i\dd x^j =\psi^4(\dd r^2+r^2\dd\Omega^2), \qquad  R(r)=\psi^2(r)r. \label{eq:conformal-spatial-metric}
\end{equation}
Unlike the homogeneous Space II-A exterior, the Space II-B initial data are radially inhomogeneous: the nontrivial profile $\psi(r)$ modifies the spatial geometry through $R(r)$, while the matter sources and trace-free extrinsic curvature may also vary with radius. The numerical Space II-B portion extends radially from the assumed virialized-core data to an outer two-sphere at $r=r_{\rm comp}$, whose areal radius is $R(r_{\rm comp})=R_{\rm comp}$. At this endpoint, the initial data close to those induced by the Space II-A FLRW exterior within the declared tolerances. For $R>R_{\rm comp}$, the same spacelike hypersurface is continued by the exact FLRW initial data induced from the analytic Space II-A spacetime.

Because the smooth construction is formulated as an ADM initial-data problem, the matter variables must be distinguished between the fluid rest frame and the Eulerian frame associated with the chosen spacelike hypersurface. We use $e$ for the rest-frame total energy density, $\varepsilon_E\equiv T_{\mu\nu}n^\mu n^\nu$ for the energy density measured by the hypersurface-normal Eulerian observer, and $p$ for pressure, all in ${\rm J\,m^{-3}}={\rm Pa}$. For a perfect fluid,
\begin{equation}
 T^{\mu\nu}=\frac{e+p}{c_0^2}u^\mu u^\nu+p g^{\mu\nu}, \label{eq:perfect-fluid-frame-decomposition}
\end{equation}
where $u^\mu$ is the fluid four-velocity and $n^\mu$ is the unit normal to the selected ADM hypersurface. Decomposing $u^\mu$ relative to $n^\mu$ with Lorentz factor $W$ gives
\begin{equation}
 \varepsilon_E=(e+p)W^2-p. \label{eq:eulerian-energy-density}
\end{equation}
Thus, $e$ and $\varepsilon_E$ coincide only when the fluid is comoving with the chosen hypersurface normal. Hereafter, the subscript $b$ on matter variables denotes their homogeneous Space II-A background values, not the GCT parameter $b$. In the homogeneous comoving background, $W=1$ and hence $\varepsilon_{E\,b}=e_b$. The corresponding Eulerian momentum density,
\begin{equation}
 j_{E\,i}=-\gamma_i{}^\mu T_{\mu\nu}n^\nu, \label{eq:eulerian-momentum-density}
\end{equation}
provides the source of the ADM momentum constraint.

Nonzero trace-free extrinsic curvature and a momentum source may therefore be present. The Hamiltonian and momentum constraints determine admissible initial data, but not lapse or shift evolution. Space II-B is one constraint-satisfying spacelike slice whose inner data represent the assumed Space III-A virialized-core sector and whose numerical endpoint returns to the Space II-A FLRW state within the declared tolerances; the analytic exterior is then set exactly to that state. It is neither a static radial interpolation between two spacetimes nor a time evolution.

\subsection{Space III-A: local proper-time and assumed virialized-core sector}
\label{sec:gct-local} 

A gravitationally bound and virialized region is described using local proper time,
\begin{equation}
 N_{\rm loc}=1, \qquad c_{\rm loc}=c_0, \qquad G_{\rm loc}=G_0. \label{eq:gct-local-sector}
\end{equation}
Here $N_{\rm loc}=1$ denotes the local proper-time representation; it does not introduce an independent regional value of the GCT parameter $b$. The usual local forms of special relativity, quantum mechanics, electromagnetism, and thermodynamics are consequently retained. No local Lorentz-violating propagation law is introduced in the construction.

The distinction is the temporal analogue of the spatial decoupling of a bound system from the Hubble flow. A virialized object does not inherit the cosmological scale factor as an expansion of its internal rulers. Operationally, its local dynamics may likewise be represented in its own proper-time normalization, $N_{\rm loc}=1$, without interpreting this as a distinct local value of the global GCT parameter $b$ or as a dynamically screened lapse field. 

The local bound sector is an operational proper-time representation, with locally calibrated constants $c_0$ and $G_0$, together with an assumed virialized-core interpretation. In the smooth construction it is represented by the inner data of the Space II-B slice; it is not an independent spacetime region. It does not establish exact or long-term staticity. It does not by itself specify a complete matter-filled halo metric. In particular, it must be distinguished from both the actual halo interior and the exterior boundary representation introduced next for the sharp calculation.

\subsection{Space III-B: local-static exterior boundary representation}
\label{sec:local-static-boundary} 

On the local side of the sharp comparison, the geometry at and outside the virial boundary is represented by the exterior Schwarzschild--de Sitter (SdS) form
\begin{equation}
 \dd s_-^2=-F_-(R)c_0^2\dd T_-^2 +\frac{\dd R^2}{F_-(R)}+R^2\dd\Omega^2, \label{eq:sharp-local-static-metric}
\end{equation}
where
\begin{equation}
 F_-(R)=1-\frac{2G_0M}{c_0^2R}-\frac{\Lambda R^2}{3}. \label{eq:fminus}
\end{equation}
Here $M$ is the mass enclosed by the boundary. This is the local-side boundary geometry for the sharp comparison, not a metric for the complete matter-filled halo interior; no use of Birkhoff's theorem inside that matter-filled region is implied.

\subsection{Sharp versus smooth: geometric roadmap}
\label{sec:geometric-roadmap} 

Thus, the two problems can be summarized without identifying their geometric objects:
\begin{equation*}
 \text{sharp:}\qquad
 \text{Space III-B}\longleftrightarrow\text{Space II-A},
\end{equation*}
\begin{equation*}
 \text{smooth:}\qquad
 \text{Space III-A core data}\subset\text{Space II-B ADM slice}
 \longrightarrow\text{Space II-A exterior}.
\end{equation*}
The sharp construction solves the Israel junction condition and finds a negative distributional surface layer on the ordinary branch. The smooth construction solves the Hamiltonian and momentum constraints from the core, through finite compensation, to the Space II-A exterior and finds a positive finite-width initial-data embedding. Figure~\ref{fig:conceptualarchitecture} shows this distinction.

This construction does not specify a radially varying interpolating profile $c(T,r)$, $G(T,r)$, or $N(T,r)$ connecting the local and cosmological representations. Nor does it derive the dynamical formation of the local normalization. Its purpose is the prior initial-data existence test: whether a positive, constraint-satisfying finite environment can replace the singular source generated by sharp matching.

\FloatBarrier

\section{Virial boundary data from spherical collapse}
\label{sec:virial} 

The spherical-collapse calculation supplies boundary data for the assumed virialized core. These quantities enter the local-static exterior boundary representation of Space III-B used in the sharp calculation and the inner source model of the Space II-B smooth initial-data construction.

The global solution supplies $a(T)$, $H_T(T)$, $c_{\rm bg}(T)$, $G_{\rm bg}(T)$, $\rho_{m,\rm bg}(T)$, and $\rho_{\Lambda,\rm bg}(T)$, where $H_T\equiv a^{-1}\dd a/\dd T$ and the corresponding background proper-time rate is $H_p=H_T/N_{\rm bg}$. These background quantities determine the turnaround and collapse environment. Once the assumed virialized-core sector is specified, however, its mechanical energies are evaluated in the local proper-time representation with the frozen constants $c_0$ and $G_0$:
\begin{equation}
 U_G=-\frac{3G_0M^2}{5R},\qquad U_\Lambda=-\frac{\Lambda c_0^2MR^2}{10}. \label{eq:virialpotentials}
\end{equation}
The vacuum term follows from the weak-field SdS potential $\Phi_\Lambda(r)=-\Lambda c_0^2r^2/6$; for a homogeneous sphere, $U_\Lambda=\int\rho\,\Phi_\Lambda\,\dd V=-\Lambda c_0^2MR^2/10$. The corrected vacuum virial theorem and energy at virialization are
\begin{equation}
 2K+U_G-2U_\Lambda=0,\qquad E_{\rm vir}=\frac{1}{2}U_G+2U_\Lambda. \label{eq:virialtheorem}
\end{equation}
The factor of two multiplying $U_\Lambda$ is fixed by the homogeneity degree of the potential, not by a fitted convention. For a homogeneous potential satisfying $U(\{\lambda\mathbf r_i\})=\lambda^nU(\{\mathbf r_i\})$, Euler's theorem gives $\sum_i\mathbf r_i\cdot\nabla_iU=nU$~\cite{Goldstein:2002}. Since the scalar virial theorem in equilibrium is $2K-\sum_i\mathbf r_i\cdot\nabla_iU=0$, a degree-$n$ potential contributes $-nU$. Newtonian self-gravity has $n=-1$ and therefore contributes $+U_G$, whereas the vacuum potential is quadratic, $n=2$, and contributes $-2U_\Lambda$.

For $w=-1$, energy conservation between turnaround ($K_{\rm ta}=0$) and virialization, together with $U_{G,\rm vir}=U_{G,\rm ta}/y_{\rm vir}$ and $U_{\Lambda,\rm vir}=U_{\Lambda,\rm ta}y_{\rm vir}^2$, leads to~\cite{Lahav:1991wc}
\begin{equation}
 4y_{\rm vir}^3 -2(\zeta Q_{\rm ta}+1)y_{\rm vir} +\zeta Q_{\rm ta}=0 \,, \label{eq:cubic}
\end{equation}
where
\begin{equation}
 y_{\rm vir}=\frac{R_{\rm vir}}{R_{\rm ta}},\qquad \zeta= \left.\frac{\rho_{\rm cl}}{\rho_{m,\rm bg}}\right|_{\rm ta},
 \qquad Q_{\rm ta}= \left.\frac{\rho_{m,\rm bg}}{\rho_{\Lambda,\rm bg}}\right|_{\rm ta}. \label{eq:yvir}
\end{equation}
Thus, $\zeta$ is the top-hat overdensity at turnaround and $Q_{\rm ta}$ is the corresponding background matter-to-vacuum density ratio. Eq.~\eqref{eq:cubic} is equivalent to the standard Lahav form
\begin{equation}
2\eta y_{\rm vir}^3-(2+\eta)y_{\rm vir}+1=0 , \quad \textrm{where} \,\, \eta=\frac{2}{\zeta Q_{\rm ta}} \label{eq:Lahav2} \,. 
\end{equation}
Thus, we obtain the $w_{\rm de}=-1$ result directly from the homogeneous vacuum potential, the corrected virial theorem, and turnaround-to-virial energy conservation, with the Lahav form above as the verified reference comparison.

$Q_{\rm ta}$ and $\zeta$ inherit background information, whereas Eq.~\eqref{eq:virialpotentials} uses $G_0,c_0$. No independent $b$-dependence is inferred unless $Q_{\rm ta}$ and $\zeta$ are rederived from the complete SI-consistent GCT collapse system.

For the reference benchmark, rather than for a fully rederived GCT collapse solution, we adopt $z_{\rm ta}=1$ in a spatially flat $\Lambda$CDM background with $\Omega_{m0}=0.3$, $\Omega_{\Lambda0}=0.7$, $w_{\rm de}=-1$, and $a_{\rm ta}=1/2$. Hence
\begin{equation}
 Q_{\rm ta}  = \frac{\Omega_{m0}}{\Omega_{\Lambda0}}a_{\rm ta}^{-3} = 3.42857.
\end{equation}
The adopted spherical-collapse solution following Refs.~\cite{Lee:2009dz,Lee:2009qq} gives $\zeta=6.46811$ and $x_{\rm vir}=a_{\rm vir}/a_{\rm ta}=1.59966$. The physical root of the corrected cubic in Eq.~\eqref{eq:cubic}, whose $w_{\rm de}=-1$ form is equivalent to the standard Lahav relation~\cite{Lahav:1991wc}, is $y_{\rm vir}=0.488485$. Therefore,
\begin{equation}
 Q_{\rm ta}=3.42857,\qquad \zeta=6.46811,\qquad y_{\rm vir}=0.488485,\qquad \Delta_{\rm vir} =
 \zeta\left(\frac{x_{\rm vir}}{y_{\rm vir}}\right)^3 = 227.147. \label{eq:virialbenchmark}
\end{equation}
These are regenerated benchmark quantities assembled from the adopted background, the reference spherical-collapse solution, the corrected virial relation, and the collapse-time solution, rather than numbers imported directly from any single reference.

These quantities provide upstream physical input rather than direct ADM boundary conditions. They set the benchmark virial and overdensity scales entering the Space III-B sharp boundary representation and the inner source normalization of the Space II-B initial-data construction. The smooth shooting conditions, $\psi(r_{\rm comp})=1$, $\mathcal A(r_{\rm comp})=0$, and $\Delta M_E(r_{\rm comp})=0$, are imposed separately. These benchmark inputs do not constitute a proof of dynamical halo formation, and the local mechanical energies used below retain the frozen constants $c_0$ and $G_0$.

\section{Sharp timelike junction and mass-compensation obstruction}
\label{sec:sharp-matching} 

\subsection{Israel junction across the timelike world tube}
\label{sec:sharp-israel} 

The sharp calculation is a timelike junction problem. It compares the local-static exterior boundary representation of Space III-B with the local-effective expanding FLRW exterior of Space II-A. The two metrics and their areal-radius invariants are defined in Eqs.~\eqref{eq:fplus-effective-form} and \eqref{eq:fminus}. The SdS side is not the matter-filled halo interior, the FLRW side is expanding rather than static, and $F_+^{\rm eff}$ is the invariant norm of the areal-radius gradient. The verified calculation is local-effective; a full timelike junction directly to the global GCT background is not derived. Appendix~\ref{app:sharp-derivation} gives the metric reductions and detailed junction-sign conventions.

Let the shell world tube have proper time $\tau$ and induced metric
\begin{equation}
\dd s_\Sigma^2 = -c_0^2\dd\tau^2 + R^2(\tau)\dd\Omega^2 . \label{eq:shell-induced-metric}
\end{equation}
For either side, its angular extrinsic curvature is
\begin{equation}
K^\theta{}_{\theta,\pm} = \frac{\epsilon_\pm}{R} \sqrt{\frac{\dot R^2}{c_0^2}+F_\pm} = \frac{\epsilon_\pm\beta_\pm}{R}, \qquad
\beta_\pm \equiv \sqrt{\frac{\dot R^2}{c_0^2}+F_\pm}, \label{eq:angular-extrinsic-curvature}
\end{equation}
where
\begin{equation}
\epsilon_\pm = \operatorname{sign} \left( n_\pm^\mu\partial_\mu R \right).
\end{equation}
An overdot denotes differentiation with respect to shell proper time.

Let the surface tensor in an orthonormal shell frame be
\begin{equation}
S^{\hat a}{}_{\hat b} = \operatorname{diag} \left( -\sigma_E,p_s,p_s \right), \qquad
\sigma_E=c_0^2\sigma_m . \label{eq:shell-surface-tensor}
\end{equation}
With $[X]=X_+-X_-$, the Israel equation
\begin{equation}
[K_{\hat a\hat b}] - h_{\hat a\hat b}[K] = -\kappa S_{\hat a\hat b}, \qquad \kappa=\frac{8\pi G_0}{c_0^4},
\end{equation}
reduces in its angular-density projection to
\begin{equation}
\epsilon_+\beta_+ - \epsilon_-\beta_- = -\frac{4\pi G_0}{c_0^4}\sigma_E R = -\frac{4\pi G_0}{c_0^2}\sigma_m R . \label{eq:israel}
\end{equation}
The density projection is sufficient for diagnosing the surface-energy obstruction considered here; no equation of state for a putative physical shell is assumed.

The ordinary branch uses the outward orientations continuously connected to the weak-field embedding,
\begin{equation}
\epsilon_+=\epsilon_-=+1.
\end{equation}
For an exactly static shell at $R=\Rvir$,
\begin{equation}
\sigma_{m,\rm vir} = -\frac{c_0^2}{4\pi G_0\Rvir} \left( \sqrt{F_+^{\rm eff}} - \sqrt{F_-} \right).
\label{eq:sigmaneg}
\end{equation}
Here ``static'' refers only to the fixed areal radius of the idealized virial boundary, $\dot R=0$; it does not imply that the exterior FLRW fluid is static or that the boundary is comoving with it.

\subsection{Positive mass excess and the negative surface layer}
\label{sec:sharp-positive-excess}  

The sign in Eq.~\eqref{eq:sigmaneg} can be established directly from the mass excess rather than inferred only from the numerical benchmarks. Using the Space II-A Friedmann relation in Eq.~\eqref{eq:friedmann} with Eq.~\eqref{eq:fplus-effective-form}, the exterior invariant may be written as
\begin{equation}
F_+^{\rm eff} = 1 - \frac{8\pi G_0}{3c_0^2} \rho_{m,\rm LE}R^2 - \frac{\Lambda R^2}{3},
\end{equation}
whereas the Space III-B invariant is given by Eq.~\eqref{eq:fminus}. Their difference is 
\begin{align}
F_+^{\rm eff}-F_- &= \frac{2G_0}{c_0^2R} \left[ M - \frac{4\pi}{3}\rho_{m,\rm LE}R^3 \right] \equiv \frac{2G_0}{c_0^2R}\, \Delta M_{\rm halo}, \label{eq:fplus-fminus-massexcess}
\end{align}
where
\begin{equation}
\Delta M_{\rm halo} \equiv M - \frac{4\pi}{3}\rho_{m,\rm LE}R^3 \label{eq:sharp-halo-excess}
\end{equation}
is the excess of the enclosed local mass over the homogeneous local-effective FLRW mass associated with the same areal sphere.

Hence, for a positive-excess halo,
\begin{equation}
\Delta M_{\rm halo}>0 \qquad\Longrightarrow\qquad F_+^{\rm eff}>F_- .
\end{equation}
On the ordinary static branch, Eq.~\eqref{eq:sigmaneg} then gives
\begin{equation}
\sigma_{m,\rm vir}<0 . \label{eq:sharp-negative-obstruction}
\end{equation}
Therefore, the negative sign is a structural consequence of imposing an exactly homogeneous Space II-A exterior immediately outside a positive-excess Space III-B boundary representation. It is not a numerical peculiarity of the particular spherical-collapse benchmarks.

The corresponding zero-shell limit makes the same mass-locking structure explicit. If $\sigma_m=0$ on the ordinary branch, Eq.~\eqref{eq:israel} gives $\beta_+=\beta_-$. The common $\dot R^2/c_0^2$ term then cancels after squaring, leaving
\begin{equation}
F_+^{\rm eff}=F_- .
\end{equation}
Using Eq.~\eqref{eq:fplus-fminus-massexcess}, this is equivalent to
\begin{equation}
\Delta M_{\rm halo}=0, \qquad\Longleftrightarrow\qquad M = \frac{4\pi}{3}\rho_{m,\rm LE}R^3 .
\label{eq:sharp-mass-locking}
\end{equation}
Thus, a shell-free sharp junction requires the local enclosed mass to equal the homogeneous FLRW matter mass associated with the same areal sphere. This is a compatibility or mass-locking condition, not a shell equation of motion and not a mechanism for virialization.

The numerical benchmarks quantify the same obstruction. The ratios in Table~\ref{tab:sharp} show that the integrated Israel surface mass is, to about $10^{-5}$ in the quoted cases, the negative of the halo excess mass. The near-equality is not expected to be exact because the Israel surface mass and the halo excess are distinct relativistic mass diagnostics. For comparison with the halo mass excess, we define the integrated Israel surface mass on the virial sphere by $M_\Sigma \equiv 4\pi R_{\rm vir}^2 \sigma_{m,\rm vir}$. The ratios in Table~\ref{tab:sharp} quantify how closely this surface mass cancels the positive halo excess in the sharp construction.

\begin{table}[!htbp]
\centering
\caption{Sharp and virial reference values for the local-effective sharp-junction calculation. The analytic sign $\sigma_m<0$ follows from Eq.~\eqref{eq:fplus-fminus-massexcess}; the listed benchmarks quantify the corresponding surface-mass cancellation.}
\begin{tabular}{lll}
\toprule
Quantity & $z_{\rm ta}=1$ & $z_{\rm ta}=2$\\
\midrule
$M_\Sigma/M$ & $-0.995608$ & $-0.994090$\\
$M_\Sigma/\Delta M_{\rm halo}$ & $-1.000010$ & $-1.000015$\\
ordinary-branch sign & negative & negative\\
\bottomrule
\end{tabular}
\label{tab:sharp}
\end{table}

\FloatBarrier

\subsection{Mass-compensation interpretation}
\label{sec:mass-compensation} 

Having first established the negative sign analytically and then quantified the near cancellation in Table~\ref{tab:sharp}, we may interpret the layer. It is neither negative halo matter nor the matter responsible for virialization. Rather, imposing exact homogeneity immediately outside a positive overdensity omits the finite environmental compensation, so the missing deficit appears as a zero-width distributional source.

The sharp layer should not be identified with a zero-thickness version of the smooth solution. It is the distributional source required when the finite compensating environment is omitted altogether.

For the smooth ADM construction, the corresponding background-relative energy-equivalent mass contrast on the conformal slice is
\begin{equation}
\Delta E_E(r) = 4\pi \int_0^r \left[ \varepsilon_E(r')-\varepsilon_{E\,b} \right] \psi^6(r')\,r'^2\,\dd r', \qquad
\Delta M_E(r) = \frac{\Delta E_E(r)}{c_0^2}, \label{eq:energy-compensation}
\end{equation}
where $\varepsilon_E$ is the Eulerian energy density defined in Sec.~\ref{sec:smooth-space-roadmap}, and $\varepsilon_{E\,b}$ is its homogeneous Space II-A background value. The factor $\psi^6$ supplies the proper-volume element on the conformal slice.

\section{Smooth positive-density initial-data embedding: Space II-B}
\label{sec:smooth} 

\subsection{Conformal ADM formulation and shooting conditions}
\label{sec:smooth-formulation} 

The smooth calculation is not a second timelike junction. As depicted on the smooth-construction side of Fig.~\ref{fig:conceptualarchitecture}, Space II-B is one spherical spacelike hypersurface extending from inner data representing the assumed Space III-A virialized core, through the finite compensating region, to the exact Space II-A local-effective FLRW exterior.

The geometric context is the general spherical ADM line element
\begin{equation}
 \dd s^2=-\alpha^2(\dd x^0)^2 +\gamma_{rr}(\dd r+\beta^r\dd x^0)^2 +R^2\dd\Omega^2 . \label{eq:smooth-adm-metric}
\end{equation}
We use the conformally flat spatial metric defined in Eq.~\eqref{eq:conformal-spatial-metric} and impose a constant-mean-curvature (CMC) condition in the standard conformal initial-data framework~\cite{York:1978gql}. The constraint equations determine the intrinsic spatial metric and extrinsic curvature on the slice; no lapse or shift evolution is solved in the present construction. The CMC condition is
\begin{equation}
 K=-\frac{3H_p}{c_0}, \label{eq:cmc}
\end{equation}
where $K$ is the trace of the extrinsic curvature and has units ${\rm m^{-1}}$ in the length-like time convention defined in Sec.~\ref{sec:gct-overview}.

The Hamiltonian and momentum constraints are
\begin{align}
 {}^{(3)}R+K^2-K_{ij}K^{ij} &=\frac{16\pi G_0}{c_0^4}\varepsilon_E+2\Lambda, \label{eq:ham}\\
 D_j(K^{ij}-\gamma^{ij}K) &=\frac{8\pi G_0}{c_0^4}j_E^i . \label{eq:mom}
\end{align}

We decompose the mixed extrinsic curvature as
\begin{equation}
 K^i{}_j = A^i{}_j+\frac{1}{3}\delta^i{}_jK, \label{eq:extrinsic-decomposition}
\end{equation}
and, for spherical symmetry, parameterize its trace-free part as
\begin{equation}
 A^i{}_j = \operatorname{diag}(2\mathcal A,-\mathcal A,-\mathcal A), \label{eq:spherical-tracefree-curvature}
\end{equation}
so that
\begin{equation}
 A_{ij}A^{ij}=6\mathcal A^2 . \label{eq:A2}
\end{equation}

The source profile contains an overdense core, a smooth infall shoulder, and a finite underdense but positive compensating region. We denote by $A_-$ the dimensionless amplitude controlling the depth of this background-relative underdensity; it is unrelated to the trace-free extrinsic-curvature variable $\mathcal A(r)$ introduced above. For a chosen underdensity amplitude $A_-<1$, the shooting conditions
\begin{equation}
 \psi(r_{\rm comp})=1,\qquad \mathcal A(r_{\rm comp})=0,\qquad \Delta M_E(r_{\rm comp})=0 \label{eq:shooting}
\end{equation}
determine the central conformal factor, velocity-profile parameter, and isotropic-coordinate endpoint $r_{\rm comp}$. Because the first condition sets $\psi=1$ there, the corresponding areal radius satisfies $R_{\rm comp}=r_{\rm comp}$.

The source construction additionally gives $\varepsilon_E=\varepsilon_{E\,b}$, $j_{E\,r}=0$, and $p=0$ at the outer endpoint, while the CMC prescription fixes $K=K_{\rm FLRW}=-3H_p/c_0$. The unused condition $\psi_{,\bar r}(\bar r_{\rm comp})\simeq0$ provides an independent geometric test of local-effective FLRW closure.

Appendix~\ref{app:spherical-constraints} gives the spherical reduction of Eqs.~\eqref{eq:ham}--\eqref{eq:mom} and the origin and endpoint conditions; the normalization and reproducibility protocol are given in Appendices~\ref{app:conventions-normalization} and \ref{app:reproducibility}. Here ``smooth'' refers to the finite-width spatial data and their finite-radius closure, not to a dynamically solved four-dimensional lapse or radial variation of $c$ or $G$.

\subsection{Positive compensated solution}
\label{sec:positive-solution} 
\suppressfloats[t]

We now present a representative positive-density compensated solution within the source family described above. The underdensity amplitude is chosen as $A_-=0.995$, while the remaining shooting parameters are determined by the conditions in Eq.~\eqref{eq:shooting}. The resulting compensation radius is solution dependent and is not a universal prediction. The representative solution gives
\begin{equation}
 A_-=0.995,\qquad \frac{\Rcomp}{\Rvir}\simeq6.2779,\qquad \min\left(\frac{e}{e_b}\right) \simeq0.00500>0 . \label{eq:best}
\end{equation}
Here $e$ is the rest-frame total energy density defined in Sec.~\ref{sec:smooth-space-roadmap}. Its local-equivalent mass density is $e/c_0^2$, so $e/e_b$ is also the corresponding local-equivalent mass-density ratio.

\begin{figure}[!htbp]
\centering
\includegraphics[width=0.90\linewidth]{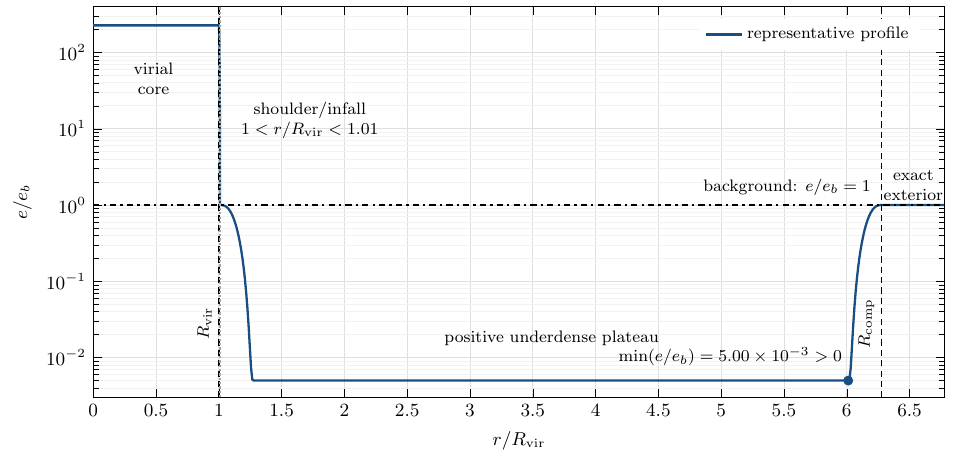}
\caption{Rest-frame energy-density ratio $e/e_b$ for the representative positive compensated solution. The inner dashed marker denotes the nominal virial/core scale $\bar r=r/R_{\rm vir}=1$, while the outer dashed marker denotes $R_{\rm comp}$; the dotted marker terminates the narrow shoulder at $\bar r=1.01$. The horizontal line denotes the homogeneous Space II-A background, $e/e_b=1$. The compensating region reaches $\min(e/e_b)\simeq5.00\times10^{-3}$ but remains strictly positive throughout. At $\Rcomp$ the numerical source returns to the homogeneous background; the complete geometric and matter closure is quantified in Sec.~\ref{sec:finite-radius-closure}. The logarithmic scale exposes both the overdense core and the positive background-relative deficit.}
\label{fig:density}
\end{figure}

Figure~\ref{fig:density} shows the morphology of the rest-frame density: an overdense core, a narrow shoulder, and a strongly underdense but positive compensation region that returns to the background at $R_{\rm comp}$. The separate WEC, NEC, and DEC audit is reported in Sec.~\ref{sec:smooth-energy-mass} and Appendix~\ref{app:locenecon}.

A simple lower bound on the required compensation radius follows from density positivity alone. As a normalization-scale estimate, the positive core excess relative to the background is proportional to $(\Delta_{\rm vir}-1)\Rvir^3$. If the compensating matter is required to remain nonnegative, the largest possible background-relative deficit per unit volume is obtained in the limiting case of vanishing local density. Consequently,
\begin{equation}
 \frac{\Rcomp}{\Rvir} \ge \Delta_{\rm vir}^{1/3} \simeq6.1015 . \label{eq:positivity-volume-bound}
\end{equation}
This is only a necessary volume bound; it does not impose the Hamiltonian or momentum constraints, the prescribed source shape, or the finite-radius geometric closure conditions. The fully coupled shooting solution instead gives $\Rcomp/\Rvir\simeq6.2779$.

The bound also shows why a finite-width compensating environment is required if the local density is to remain nonnegative. Since the background-relative deficit per unit volume is bounded by density positivity, the positive core excess cannot be compensated in an arbitrarily thin layer outside the nominal core/profile scale. The fact that the coupled solution lies only modestly above the positivity bound is a property of the representative source family; it does not imply that the physical compensation profile or its detailed microphysics is unique.

\begin{figure}[!htbp]
\centering
\includegraphics[width=0.90\linewidth]{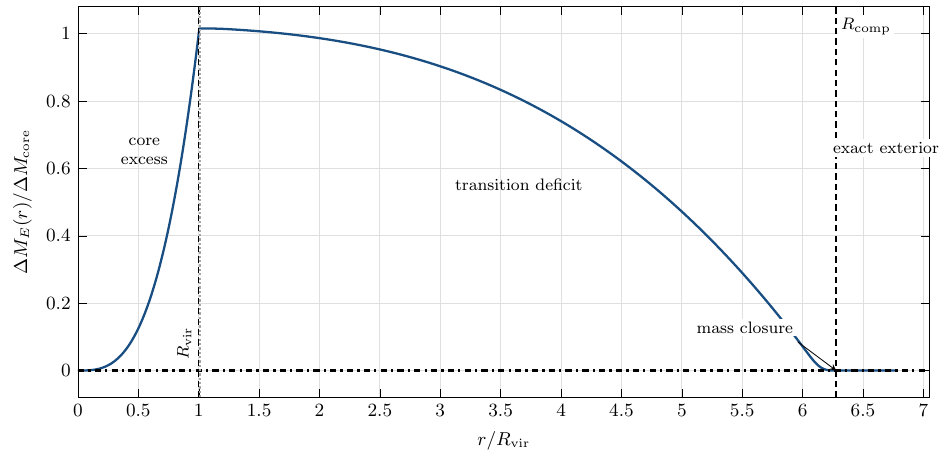}
\caption{Cumulative proper-volume energy-equivalent mass contrast, $\Delta M_E(r)/\Delta M_{\rm core}$, for the representative slice. The overdense core builds positive cumulative excess, while the finite-width underdense compensation region progressively removes it. The curve returns to zero at $\Rcomp$ and remains zero in the exact analytic Space II-A exterior. Markers denote the nominal virial/core scale $\bar r=1$, the shoulder endpoint $\bar r=1.01$, and the compensation endpoint $R_{\rm comp}$. This is a slice-dependent integral mass-closure diagnostic and should not be identified with either the Israel surface mass or the Misner--Sharp mass length.}
\label{fig:compensation}
\end{figure}

The cumulative energy-equivalent mass contrast defined in Eq.~\eqref{eq:energy-compensation} uses the proper-volume element $\psi^6r^2\dd r\,\dd\Omega$. Figure~\ref{fig:compensation} shows its radial closure.

\FloatBarrier

The endpoint condition
\begin{equation}
 \Delta M_E(r_{\rm comp})=0
\end{equation}
represents integral mass closure on the chosen ADM slice, rather than merely pointwise recovery of the background density. Beyond $R_{\rm comp}$ the exact Space II-A exterior has zero background-relative contrast, so the cumulative quantity remains zero. This slice-dependent diagnostic is distinct from both the Israel surface mass and the invariant Misner--Sharp mass length.

Numerically,
\begin{align}
&\Delta\bar M_{\rm core}
 =+2.01705565\times10^{-5},&
\Delta\bar M_{\rm trans}
 &=-2.01705565\times10^{-5}, \nonumber \\
&\overline{\Delta M}_E(\bar r_{\rm comp})
 =1.6383\times10^{-20}. \label{eq:smoothmassclosure}
\end{align}

Therefore, the positive core excess and the finite-width compensating deficit cancel to the numerical precision of the representative coupled solution.

\FloatBarrier

\subsection{Finite-radius FLRW closure}
\label{sec:finite-radius-closure} 

The three solved shooting conditions are those in Eq.~\eqref{eq:shooting}: $\psi(r_{\rm comp})=1$, $\mathcal A(r_{\rm comp})=0$, and $\Delta M_E(r_{\rm comp})=0$. Separately, the source construction gives $\varepsilon_E=\varepsilon_{E\,b}$, $j_{E\,r}=0$, and $p=0$ at the endpoint, while the CMC prescription gives $K=K_{\rm FLRW}=-3H_p/c_0$. The unused condition $\psi_{,\bar r}(\bar r_{\rm comp})\simeq0$ is an independent closure diagnostic, not a fourth shooting condition.

The representative endpoint residuals are
\begin{equation}
\psi-1=2.22\times10^{-16},\quad \psi_{,\bar r}=2.47\times10^{-12},\quad \bar{\mathcal A}=-1.85\times10^{-26}, \label{eq:matchres}
\end{equation}
with the source and CMC endpoint values above. Beyond the numerical endpoint, the exterior is the exact analytic Space II-A solution. The constructed initial slice therefore has no residual distributional layer; this statement is not a theorem about an evolved timelike junction.

Density matching alone is insufficient: the conformal geometry, its radial derivative, the trace-free and mean extrinsic curvature, and the momentum source also return to their Space II-A values within the declared tolerances. Detailed endpoint profiles are given in Appendix~\ref{app:finradFLRW}.

\subsection{Energy conditions and mass diagnostics}
\label{sec:smooth-energy-mass} 

For the isotropic perfect-fluid source defined in Sec.~\ref{sec:smooth-space-roadmap}, the stress tensor is diagonal in its local orthonormal rest frame,
\begin{equation}
 T^{\hat a}{}_{\hat b}=\operatorname{diag}(-e,p,p,p), \label{eq:perfect-fluid-rest-frame}
\end{equation}
with both $e$ and $p$ in ${\rm J\,m^{-3}}$. The standard pointwise energy-condition inequalities are~\cite{Hawking:1973uf}
\begin{equation}
 \mathrm{WEC}: e\ge0,\ e+p\ge0;\qquad \mathrm{NEC}: e+p\ge0;\qquad \mathrm{DEC}: e\ge|p|. \label{eq:ecs}
\end{equation}
They hold for the representative solution in dimensionless ratios, which are invariant under the common positive SI energy-density scaling.

The three tested margins satisfy
\begin{equation}
\min e=\min(e+p)=\min(e-|p|) =1.06\times10^{-10} \label{eq:ecmin}
\end{equation}
in barred energy-density units.  WEC, NEC, and DEC are strictly satisfied; SEC is not evaluated.  

Thus, under the adopted isotropic perfect-fluid closure, removing the negative distributional layer does not require exotic bulk matter. This is an initial-slice matter-admissibility result; the strong energy condition is not tested. Appendix~\ref{app:locenecon} gives the numerical margin audit.

The endpoint invariant mass-length contrast~\cite{Misner:1964je} is
\begin{equation}
\Delta\bar\mu_{\rm MS}\simeq-1.94\times10^{-10}, \label{eq:msres}
\end{equation}
distinct from $\overline{\Delta M}_E$ and within the declared $10^{-9}R_{\rm vir}$ tolerance. Unlike the slice-dependent proper-volume contrast $\Delta M_E$, $\mu_{\rm MS}$ is reconstructed from the spherical spacetime geometry; agreement is required only at the closure endpoint, not pointwise at intermediate radii. The derivative sensitivity and invariant reconstruction are detailed in Appendix~\ref{app:invmisshr}.

The corresponding barred masses are
\begin{equation}
\Delta\bar M_{\rm core}=+2.01706\times10^{-5},\quad
\Delta\bar M_{\rm trans}=-2.01706\times10^{-5},\quad
\bar M_\Sigma=-2.01696\times10^{-5}. \label{eq:masscompare}
\end{equation}
The sharp and smooth negative masses are nearly, but not exactly, equal because an Israel surface mass and a proper-volume contrast are different relativistic diagnostics.

Table~\ref{tab:pp2} collects the representative solution parameters and the principal profile, constraint, endpoint, and mass-closure diagnostics. Their definitions and full residual profiles are given in Appendix~\ref{app:admissibility}.

\begin{table}[!htbp]
\centering
\caption{Representative positive-compensation parameters and diagnostics.}
\begin{tabular}{ll}
\toprule
Quantity & Value\\
\midrule
$A_-$ & $0.995$\\
$\Rcomp/\Rvir$ & $6.277893$\\
$\min(e/e_b)$ & $0.00499994$\\
$\|\bar{\mathcal H}\|_\infty$ & $2.6316\times10^{-8}$\\
$\|\bar{\mathcal M}_r\|_\infty$ & $1.3646\times10^{-11}$\\
$\psi_{,\bar r}(\bar r_{\rm comp})$ & $2.4672\times10^{-12}$\\
$\overline{\Delta M}_E(\bar r_{\rm comp})$ & $1.6383\times10^{-20}$\\
\bottomrule
\end{tabular}
\label{tab:pp2}
\end{table}

Together, these checks establish finite-radius geometric and matter closure without negative local energy density. They do not identify the slice-dependent $\Delta M_E$, the Israel surface mass $M_\Sigma$, and the invariant mass length $\mu_{\rm MS}$: their numerical near-equalities are diagnostic relations, not exact identities.

\FloatBarrier


\section{Numerical validation, discussion, and claim boundaries}
\label{sec:validation} 

\subsection{Independent numerical validation}
\label{sec:independent-validation} 

Independent manufactured-state and null-background tests validate the spherical spatial, constraint, origin-regularization, and mass-diagnostic operators. Table~\ref{tab:constraints} reports the multi-resolution constraint convergence; the test protocol and detailed orders are given in Appendix~\ref{app:reproducibility}.

\FloatBarrier
\begin{table}[!htbp]
\centering
\caption{Constraint convergence for the representative solutions in the barred normalization. At every $N$, the complete shooting problem is solved again before the independent residual norms are evaluated.}
\begin{tabular}{rcc}
\toprule
$N$&$\|\bar{\mathcal H}\|_\infty$&
$\|\bar{\mathcal M}_r\|_\infty$\\
\midrule
200&$1.71444\times10^{-6}$&$2.13563\times10^{-10}$\\
400&$1.22112\times10^{-7}$&$5.44093\times10^{-11}$\\
800&$2.63164\times10^{-8}$&$1.36455\times10^{-11}$\\
1600&$6.72375\times10^{-9}$&$3.49652\times10^{-12}$\\
\bottomrule
\end{tabular}
\label{tab:constraints}
\end{table}

These checks support the numerical map from the prescribed sources to the reconstructed initial slice. As specified in Sec.~\ref{sec:claim-boundaries}, they are spatial initial-data tests rather than tests of dynamical formation, stability, or persistence.

\FloatBarrier

\subsection{Physical interpretation and relation of the two constructions}
\label{sec:discussion} 

The two constructions answer successive but geometrically distinct questions. In the sharp Space III-B--Space II-A junction, a positive excess placed immediately next to a homogeneous exterior requires a negative distributional Israel layer. In the smooth construction, Space III-A core data lie on one Space II-B spacelike slice; a finite-width, background-relative deficit with positive local density permits constraint-satisfying closure to Space II-A. Thus, the smooth construction is not a replacement timelike shell, and the sharp layer is not exotic halo matter but may be interpreted as the distributional image of the omitted environmental compensation.

\subsection{Precise claim boundaries}
\label{sec:claim-boundaries} 

The result establishes initial-data existence and admissibility only, within the adopted source family, benchmark, and closure prescription. It proves neither dynamical formation nor stability or persistence. It does not supply a full timelike junction to the global GCT background, a radial $N(T,r)$, $c(T,r)$, or $G(T,r)$ interpolation, or dynamical screening of $b$; no Vainshtein- or chameleon-type mechanism is introduced. No uniqueness theorem is claimed, and no assertion is made that every virialized halo or source family admits the same embedding.

\section{Conclusions}
\label{sec:conclusions} 

The purpose of this work was to determine whether a positive-excess virialized local sector can be embedded consistently in a homogeneous $\Lambda$--FLRW environment once the idealized zero-width junction used in the sharp construction is replaced by a finite environmental transition. The two calculations considered here address this question from complementary geometric viewpoints.

For the ordinary sharp Space III-B--Space II-A timelike matching, the mass-excess relation shows analytically that $\Delta M_{\rm halo}>0$ implies $F^{\rm eff}_+>F_-$ and therefore $\sigma_m<0$. Thus, the negative Israel layer is not a peculiarity of the numerical benchmark but a structural consequence of placing an exactly homogeneous FLRW exterior immediately outside a positive mass excess. Conversely, the zero-shell limit enforces the mass-locking condition $M=(4\pi/3)\rho_{m,\rm LE}R^3$ and does not provide an independent mechanism for representing a virialized overdensity. The sharp obstruction is consequently interpreted as the distributional representation of the finite environmental compensation that has been omitted by the zero-width idealization.

The smooth construction shows that this obstruction need not persist once the environmental transition is resolved over a finite radial region. Starting from inner data representing an assumed virialized Space III-A core, we constructed a conformally flat CMC initial hypersurface whose finite-width compensating region remains everywhere positive in the local rest frame and closes at finite radius to the exact analytic Space II-A FLRW exterior. The representative solution satisfies the Hamiltonian and momentum constraints under resolution refinement, returns its geometric, extrinsic-curvature, and matter variables to the FLRW endpoint values within the stated tolerances, satisfies the tested weak, null, and dominant energy conditions, and passes both the proper-volume mass-compensation and invariant Misner--Sharp endpoint diagnostics. Thus, the negative distributional layer of the sharp construction can be replaced, on a single spacelike initial hypersurface, by a regular finite-width positive-density environment.

The physical implication is that an assumed virialized local proper-time sector and a homogeneous expanding FLRW exterior are not incompatible at the level of the Einstein constraint equations. Within the adopted source family and closure prescription, there exist regular boundary and initial-data conditions that connect the bound core through a compensating environment to the homogeneous exterior without leaving a residual thin shell on the constructed slice. This is the principal constructive result of the present work.

In the GCT interpretation, this result removes a specific geometric obstruction to the coexistence of the global cosmological clock sector and the locally bound proper-time sector. Space II-A is the homogeneous proper-time, local-effective representation of the Space I background, while the inner data represent the assumed local proper-time sector with the reference constants $c_0$ and $G_0$ held fixed. The existence of a positive, constraint-satisfying finite-width embedding therefore demonstrates that a necessary geometric initial-data prerequisite for this local--global clock interpretation can be satisfied. It does not, however, by itself establish the dynamical formation of that sector or a four-dimensional transition between the two clock descriptions.

Accordingly, the present result is a constructive initial-data existence and admissibility demonstration, not a dynamical screening or stability theorem. A full dynamical realization would require evolving such initial data from a pre-virial cosmological overdensity, determining whether the local proper-time sector forms and persists dynamically, and constructing, if required, the corresponding inhomogeneous spacetime dependence of $N(T,r)$ and the associated GCT quantities. Those questions constitute the next stage beyond the geometric existence result established here.

\FloatBarrier

\section*{Acknowledgments}

This work is supported by the Basic Science Research Program through the National Research Foundation of Korea (NRF), funded by the Ministry of Science and ICT under Grant No. NRF2022R1A2C1005050. During the preparation of this work, the author used ChatGPT (OpenAI) as an interactive assistant for literature organization, manuscript and \LaTeX{} preparation, improving the clarity and readability of the English text, and coding and figure-workflow support. The author directed, checked, and revised all AI-assisted outputs used in this work. The research ideas, scientific reasoning, assessment of the physical and mathematical validity of the analysis, equations, numerical methodology, interpretation of the results, and scientific validation remain the responsibility of the author, who takes full responsibility for the content and conclusions of this work.

\appendix
\renewcommand{\thefigure}{A\arabic{figure}}
\renewcommand{\theHfigure}{A.\arabic{figure}}
\setcounter{figure}{0}

\section{Conventions, SI dimensions, and numerical normalization}
\label{app:conventions-normalization}

\subsection{Symbols, meanings, and SI units}
\label{sisym} 

\footnotesize
\begin{tabular}{@{}
p{2.6cm}
p{7.0cm}
p{1.9cm}
p{2.5cm}
p{3.0cm}@{}}
\toprule
Symbol & Meaning & SI unit & Sector & Role\\
\midrule
$T$
& adopted cosmological time coordinate
& s & common & physical\\

$N_{\rm bg},N_{\rm loc}$
& background and local lapse normalizations
& 1 & bg/local & physical convention\\

$c_{\rm bg},c_0$
& background clock coefficient and reference/local speed
& m/s & bg/local & physical\\

$G_{\rm bg},G_0$
& background and reference/local Newton couplings
& m$^3$/kg/s$^{2}$ & bg/local & physical\\

$r,R,\Rvir,\Rcomp$
& coordinate and areal radii
& m & local & physical\\

$a,\alpha,\psi,A_-,b$
& dimensionless variables
& 1 & stated context & physical or diagnostic\\

$\mathcal A=A^r{}_r/2$
& physical mixed trace-free curvature amplitude
& 1/m & local & physical\\

$\rho_{m,\rm bg},\rho_{\Lambda,\rm bg}$
& global background mass densities
& kg/m$^{3}$ & bg & physical\\

$\rho_{m,\rm LE},\rho_{\Lambda,\rm LE}$
& $c_0^2$-calibrated mass-density forms
& kg/m$^{3}$ & mapping & conversion\\

$e,e_b$
& rest-frame total and homogeneous-background energy densities
& J/m$^{3}$ & local/bg & physical\\

$\varepsilon_E,\varepsilon_{E\,b}$
& Eulerian energy density and its comoving background value
& J/m$^{3}$ & local/bg & physical\\

$Q(a),Q_{\rm ta}$
& calibration-independent matter-to-vacuum ratio
& 1 & bg/mapping & diagnostic\\

$\zeta$
& turnaround cluster overdensity
& 1 & bg/mapping & diagnostic\\

$F_+^{\rm eff},F_+^{\rm GCT}$
& local-effective exterior norm and candidate global norm
& 1 & local/bg & diagnostic\\

$H_p=a^{-1}\dd a/\dd T_+$
& Space II-A proper-time Hubble rate
& 1/s & bg & physical\\

$\Lambda$
& cosmological constant
& 1/m$^{2}$ & common & physical\\

$p$
& pressure
& J/m$^{3}$ & local & physical\\

$j_{E\,r},j_E^{\hat r}$
& covariant and orthonormal radial momentum components
& J/m$^{3}$ & local & physical\\

$K_{ij},K$
& extrinsic curvature for $x^0=c_0T_+$
& 1/m & local & physical\\

$\sigma_m,\sigma_E$
& surface mass and surface energy densities
& kg/m$^{2}$, J/m$^{2}$ & shell & physical\\

$\mu_{\rm MS}$
& invariant mass length
& m & common & diagnostic\\

$m_{\rm MS},M$
& physical masses
& kg & stated sector & physical\\
\bottomrule
\end{tabular}
\normalsize

\FloatBarrier

\subsection{Numerical normalization and SI restoration}
\label{code} 

Choose a fixed local length $R_*=\Rvir$ and define
\begin{gather}
 \bar r=\frac{r}{R_*},\qquad
 \bar x^0\equiv\frac{x^0}{R_*} =\frac{c_0T_+}{R_*},\qquad
 \bar K=R_*K,\qquad
 \bar\Lambda=\Lambda R_*^2, \nonumber \\
 \bar e=\frac{G_0R_*^2}{c_0^4}e,\quad
 \bar\ee_E=\frac{G_0R_*^2}{c_0^4}\ee_E,\quad
 \bar p=\frac{G_0R_*^2}{c_0^4}p,\quad
 \bar\mu_{\rm MS}=\frac{\mu_{\rm MS}}{R_*} =\frac{G_0m_{\rm MS}^{\rm loc}}{c_0^2R_*}.
\label{eq:bars}
\end{gather}
Only after these definitions may a code use unity-valued reference
constants.  Restoration is obtained by inverting Eq.~\eqref{eq:bars}.
In the global background, a mass in kilograms inferred from the same
invariant length uses $c_{\rm bg}^2(T)/G_{\rm bg}(T)$, not
$c_0^2/G_0$.

\FloatBarrier

\section{Analytic derivations for the virial and sharp constructions}
\label{app:sharp-derivation} 

\subsection{\texorpdfstring{$\Lambda$}{Lambda} virial relation}
\label{app:Lvirrel} 

For a uniform sphere the local SdS acceleration is $\ddot R=-G_0M/R^2+\Lambda c_0^2R/3$. Integrating the vacuum potential over the sphere gives Eq.~\eqref{eq:virialpotentials}. Because the vacuum potential is homogeneous of degree two, its virial contribution is $-2U_\Lambda$. Applying energy conservation at turnaround $(K_{\rm ta}=0)$, Eq.~\eqref{eq:virialtheorem} at virialization, and the definitions of $y_{\rm vir},Q_{\rm ta},\zeta$ yields Eq.~\eqref{eq:cubic}. The factor of two in this argument is the difference between the corrected and earlier benchmark roots.

For the sharp-junction derivation, we define $[X]=X_+-X_-$ and use the orientation $\epsilon_\pm=\operatorname{sign}(n_\pm^\mu\partial_\mu R)$.

\subsection{Space III-B local-static boundary side}
\label{app:IIIBlocstabou} 

Using the length-like time coordinate $x_-^0 \equiv c_0 T_-$, the local-side metric in Eq.~\eqref{eq:sharp-local-static-metric} may be written as
\begin{equation}
 ds_-^2 = -F_-(R)(dx_-^0)^2 +\frac{dR^2}{F_-(R)}+R^2d\Omega^2 . \label{eq:local-static-metric-length-coordinate}
\end{equation}
The shell trajectory is then
\begin{equation}
 x_-^\mu(\tau) = \bigl(c_0T_-(\tau),R(\tau),\theta,\phi\bigr), \qquad
 u_-^\mu \equiv \frac{dx_-^\mu}{d\tau} = \bigl(c_0\dot T_-,\dot R,0,0\bigr) \,, \label{eq:local-shell-trajectory}
\end{equation}
where an overdot denotes differentiation with respect to the shell proper time \(\tau\).

The normalization $g_{\mu\nu}u_-^\mu u_-^\nu=-c_0^2$ gives
\begin{equation}
 -F_-c_0^2\dot T_-^2 +\frac{\dot R^2}{F_-} = -c_0^2, \qquad
 F_-\dot T_- \equiv \beta_- = \sqrt{F_-+\frac{\dot R^2}{c_0^2}}, \label{eq:local-velocity-normalization}
\end{equation}
for the future-directed branch. A unit normal covector is
\begin{equation}
 n^-_\mu = \epsilon_- \left( -\frac{\dot R}{c_0}, \dot T_-, 0, 0 \right). \label{eq:local-unit-normal}
\end{equation}
Direct substitution verifies
\begin{equation}
 n^-_\mu u_-^\mu=0, \qquad n^-_\mu n_-^\mu=1.
\end{equation}
Moreover,
\begin{equation}
 n_-^\mu\partial_\mu R = n_-^R = \epsilon_-F_-\dot T_- = \epsilon_-\beta_- .
\end{equation}
Since \(g_{\theta\theta}=R^2\), the angular component of the extrinsic curvature is
\begin{equation}
 K^\theta{}_{\theta,-} = \frac{1}{2}g^{\theta\theta} n_-^a\partial_a g_{\theta\theta} = \frac{n_-^a\partial_aR}{R} = \frac{\epsilon_-\beta_-}{R},
\label{eq:local-angular-curvature-derivation}
\end{equation}
where \(a\in\{x_-^0,R\}\) labels the two-dimensional orbit space. The SdS metric is used here only as the exterior boundary geometry associated with the enclosed mass $M$; it is not taken as a metric for the matter-filled halo interior.

\subsection{Space II-A local-effective FLRW side}
\label{app:IIAloceffFLRW} 

Again, by using the length-like time coordinate $x_+^0 \equiv c_0 T_+$ (\textit{i.e.}, $dx_+^0=c_0\,dT_+$), the areal-radius form of the local-effective FLRW metric, obtained from Eqs.~\eqref{eq:local-effective-flrw-metric} and~\eqref{eq:areal-radius-transform}, becomes
\begin{equation}
 ds_+^2 = -(dx_+^0)^2 + \left( dR-\frac{H_pR}{c_0}dx_+^0 \right)^2 + R^2d\Omega^2 . \label{eq:flrw-areal-radius-length-coordinate}
\end{equation}
Equivalently,
\begin{equation}
 ds_+^2 = -\left(1-\frac{H_p^2R^2}{c_0^2} \right)(dx_+^0)^2 -2\frac{H_pR}{c_0}\,dx_+^0dR +dR^2 +R^2d\Omega^2 .
\end{equation}

Its $(x_+^0,R)$ metric block and inverse are
\begin{equation}
 g_{ab} = \begin{pmatrix} -\left(1-\dfrac{H_p^2R^2}{c_0^2}\right) &-\dfrac{H_pR}{c_0} \\[6pt]
 -\dfrac{H_pR}{c_0} & 1 \end{pmatrix},
 \qquad
 g^{ab} = \begin{pmatrix} -1 & -\dfrac{H_pR}{c_0} \\[6pt]
 -\dfrac{H_pR}{c_0} & 1-\dfrac{H_p^2R^2}{c_0^2} \end{pmatrix}.\label{eq:flrw-radial-blocks}
\end{equation}
Here \(a,b\in\{x_+^0,R\}\), and all displayed metric components are dimensionless.

Because $R$ is a coordinate in this chart,
\begin{equation}
 g_+^{ab}\partial_aR\,\partial_bR = g_+^{RR} = 1-\frac{H_p^2R^2}{c_0^2} \equiv F_+^{\rm eff}. \label{eq:fplus-invariant-derivation}
\end{equation}
This scalar is the coordinate-independent spherical norm of the areal-radius gradient. No time-independent or diagonal static FLRW geometry is inferred from it.

For a shell trajectory on this side,
\begin{equation}
 x_+^\mu(\tau) = \bigl(c_0T_+(\tau),R(\tau),\theta,\phi\bigr), \qquad
 u_+^\mu = \bigl(c_0\dot T_+,\dot R,0,0\bigr),
\end{equation}
decompose the areal-radius gradient in the orthonormal pair formed by the shell four-velocity and its unit normal. In the two-dimensional orbit space,
\begin{equation}
 g^{ab} = -\frac{u_+^a u_+^b}{c_0^2} +n_+^a n_+^b .
\end{equation}
Therefore,
\begin{equation}
 F_+^{\rm eff} = -\frac{\left(u_+^a\partial_aR\right)^2}{c_0^2} + \left(n_+^a\partial_aR\right)^2 .\label{eq:flrw-gradient-decomposition}
\end{equation}
Since
\begin{equation}
 u_+^a\partial_aR=\dot R,
\end{equation}
the orientation convention
$\epsilon_+=\operatorname{sign}(n_+^a\partial_aR)$ gives
\begin{equation}
 n_+^a\partial_aR = \epsilon_+ \sqrt{\frac{\dot R^2}{c_0^2} + F_+^{\rm eff}} \equiv \epsilon_+\beta_+, 
 \qquad  \beta_+ \equiv \sqrt{\frac{\dot R^2}{c_0^2}+ F_+^{\rm eff}} .
\end{equation}
Since \(g_{\theta\theta}=R^2\), the angular extrinsic curvature is
\begin{equation}
 K^\theta{}_{\theta,+} = \frac{1}{2}g^{\theta\theta} n_+^a\partial_a g_{\theta\theta} = \frac{n_+^a\partial_aR}{R} =  \frac{\epsilon_+\beta_+}{R}.
\label{eq:flrw-angular-curvature-derivation}
\end{equation}

\subsection{Israel projection and zero-shell limit}
\label{app:Isrpro} 

The indices $a,b,\ldots$ denote intrinsic coordinates on the shell hypersurface $\Sigma$, whereas hatted indices $\hat a,\hat b,\ldots$ denote components in an orthonormal frame intrinsic to $\Sigma$. The intrinsic coordinates on the timelike junction hypersurface $\Sigma$ are chosen as
\begin{equation}
 y^a=(y^0,\theta,\phi) = (c_0\tau,\theta,\phi).
\end{equation}
Thus, the induced line element is 
\begin{equation}
 ds_\Sigma^2 = -(dy^0)^2 +R^2(\tau)d\theta^2 +R^2(\tau)\sin^2\theta\,d\phi^2,
\end{equation}
so that
\begin{equation}
 h_{ab} = \operatorname{diag} \left( -1,R^2,R^2\sin^2\theta \right).
\end{equation}
The corresponding timelike tangent basis vector is
\begin{equation}
 e^\mu{}_0 = \frac{\partial x^\mu}{\partial y^0} = \frac{u^\mu}{c_0},
\end{equation}
and satisfies $g_{\mu\nu}e^\mu{}_0e^\nu{}_0=-1$. In the associated orthonormal shell frame,
\begin{equation}
 h_{\hat a\hat b} = \operatorname{diag}(-1,1,1).
\end{equation}

In an orthonormal shell frame, the surface tensor is
\begin{equation}
 S^{\hat a}{}_{\hat b} = \operatorname{diag}(-\sigma_E,p_s,p_s) =\operatorname{diag}(-c_0^2\sigma_m,p_s,p_s).
\label{eq:appendix-surface-tensor}
\end{equation}
Substitution into
\begin{equation}
 [K_{\hat a\hat b}]-h_{\hat a\hat b}[K] = -\frac{8\pi G_0}{c_0^4}S_{\hat a\hat b} \label{eq:appendix-israel-tensor}
\end{equation}
shows from the proper-time projection that $[K^\theta{}_{\theta}]=-4\pi G_0\sigma_E/c_0^4$. Equations~\eqref{eq:local-angular-curvature-derivation} and \eqref{eq:flrw-angular-curvature-derivation} then give the unsquared master equation~\eqref{eq:israel}. On the ordinary branch $\epsilon_+=\epsilon_-=+1$ and in the static limit $\dot R=0$, it reduces directly to Eq.~\eqref{eq:sigmaneg}. Therefore, the verified ordering $F_+^{\rm eff}>F_-$ fixes the negative sign without squaring the junction equation.

Because $\sigma_E=c_0^2\sigma_m$, the result $\sigma_m<0$ is equivalently $\sigma_E<0$. Thus, on the ordinary branch, an exact zero-width junction between the local-static boundary geometry and the local-effective FLRW exterior requires a shell with negative surface energy density. This is the sharp thin-shell obstruction identified in Sec.~\ref{sec:sharp-matching}. It should not be interpreted as evidence for a physical exotic-matter layer in the halo. Rather, within the present construction, it indicates that the finite-width compensating region omitted by the sharp idealization has been compressed into a distributional surface source.

If $\sigma_m=0$, the same unsquared equation gives $\epsilon_+\beta_+=\epsilon_-\beta_-$. On the ordinary branch the common $\dot R^2/c_0^2$ term cancels after squaring, leaving $F_+=F_-$. This is a sharp-matching compatibility or mass-locking condition, not an equation of motion and not a virialization mechanism. Reversing a normal reverses its $\epsilon_\pm$ consistently and cannot be used to change only the reported matter sign.

\FloatBarrier

\section{Spherical conformal initial-data equations and boundary conditions}
\label{app:spherical-constraints} 

This appendix derives the spherical initial-data equations used in the smooth construction from the ADM constraints of Sec.~\ref{sec:smooth-formulation}. We use throughout the same Einstein--$\Lambda$ equations and extrinsic-curvature sign convention as in Sec.~\ref{sec:smooth}. The derivation is first written in terms of the physical isotropic radius $r$ and is then converted to the dimensionless coordinate $\bar r=r/R_*$, with $R_*=\Rvir$, for numerical integration.

\subsection{Spatial curvature and extrinsic curvature}
\label{app:spatial-curvature} 

For the conformally flat spatial metric in Eq.~\eqref{eq:conformal-spatial-metric}, the three-dimensional conformal transformation law gives
\begin{equation}
 {}^{(3)}R = \psi^{-4} \left[ -8\psi^{-1}\nabla_{\rm flat}^2\psi \right] = -8\psi^{-5} \left( \psi_{,rr} +\frac{2}{r}\psi_{,r}  \right), \label{eq:conformal-scalar-curvature}
\end{equation}
where the derivatives are with respect to the physical isotropic radius $r$. Using the decomposition of the mixed extrinsic curvature given in Eq.~\eqref{eq:extrinsic-decomposition} together with the spherical trace-free form introduced in Eq.~\eqref{eq:spherical-tracefree-curvature}, one finds $A_{ij}A^{ij}=6\mathcal A^2$ in Eq.~\eqref{eq:A2}. Hence, one obtains
\begin{equation}
 K_{ij}K^{ij} = 6\mathcal A^2+\frac{1}{3}K^2, \label{eq:extrinsic-curvature-contraction}
\end{equation}
where $\mathcal A$ is the physical mixed-component amplitude of the trace-free extrinsic curvature, while $K$ is its trace; both have dimensions of inverse length. Their dimensionless counterparts are $\bar{\mathcal A}=R_*\mathcal A$ and $\bar K=R_*K$.

\subsection{Hamiltonian and momentum ODE reduction}
\label{HammomODE} 

For the numerical construction, we set $R_*=\Rvir$ and $\bar r \equiv r/R_*$ as in Eq.~\eqref{eq:bars}. Here $R_{\rm vir}$ is the physical reference scale used to normalize the isotropic coordinate $r$, whereas the areal radius on the conformal slice is $R(r)=\psi^2(r)r$ as in Eq.~\eqref{eq:conformal-spatial-metric}. The prescribed source profile places the nominal virial/core scale at $\bar r=1$ and connects it to the outer compensating region through the finite-width shoulder $1<\bar r<1.01$. Since $\psi$ need not equal unity at $\bar r=1$, this coordinate location need not coincide exactly with the areal-radius condition $R=R_{\rm vir}$. For the representative solution the difference is only of order $10^{-5}$ in $R/R_{\rm vir}$, much smaller than the prescribed shoulder width $\Delta\bar r=0.01$. Derivatives with respect to the physical and dimensionless radial coordinates are related by
\begin{equation}
 \frac{\partial}{\partial r} = \frac{1}{R_*} \frac{\partial}{\partial\bar r}, \qquad \psi_{,r} = \frac{1}{R_*}\psi_{,\bar r}, \qquad
 \psi_{,rr} = \frac{1}{R_*^2}\psi_{,\bar r\bar r}. \label{eq:radial-derivative-rescaling}
\end{equation}

One can rewrite the Hamiltonian constraint in Eq.~\eqref{eq:ham} by using Eqs.~\eqref{eq:conformal-scalar-curvature} and \eqref{eq:extrinsic-curvature-contraction} to obtain
\begin{equation}
 -8\psi^{-5} \left( \psi_{,rr} +\frac{2}{r}\psi_{,r} \right) +\frac{2}{3}K^2 -6\mathcal A^2  = \frac{16\pi G_0}{c_0^4}\varepsilon_E  +2\Lambda. \label{eq:physical-spherical-hamiltonian}
\end{equation}
The exact homogeneous exterior data induced from Space II-A provide the reference solution for the subtraction. In the spatial normalization adopted for the flat-FLRW exterior,
\begin{equation}
 \psi_{\rm FLRW}=1, \qquad \psi_{{\rm FLRW},r}=0, \qquad {}^{(3)}R_{\rm FLRW}=0, \qquad \mathcal A_{\rm FLRW}=0, \qquad \varepsilon_E=\varepsilon_{E\,b}, \qquad
 K_{\rm FLRW}=-\frac{3H_p}{c_0}.
\end{equation}
Then, the Hamiltonian constraint reduces to
\begin{equation}
 \frac{2}{3}K_{\rm FLRW}^2 = \frac{16\pi G_0}{c_0^4}\varepsilon_{E\,b} +2\Lambda. \label{eq:physical-background-hamiltonian}
\end{equation}
Equivalently, using the barred variables of Eq.~\eqref{eq:bars},
\begin{equation}
 \frac{2}{3}\bar K_{\rm FLRW}^{\,2} = 16\pi\bar\varepsilon_{E\,b} +2\bar\Lambda. \label{eq:barred-background-hamiltonian}
\end{equation}
Thus, Eq.~\eqref{eq:barred-background-hamiltonian} is simply the Hamiltonian-constraint form of the homogeneous Space II-A flat-FLRW background relation.

The CMC construction fixes $K$ to this same FLRW value throughout the initial slice. Subtracting Eq.~\eqref{eq:physical-background-hamiltonian} from Eq.~\eqref{eq:physical-spherical-hamiltonian} therefore removes both the CMC term and the cosmological-constant term and gives
\begin{equation}
 -8\psi^{-5} \left( \psi_{,rr} +\frac{2}{r}\psi_{,r} \right) -6\mathcal A^2 = \frac{16\pi G_0}{c_0^4} \left( \varepsilon_E-\varepsilon_{E\,b} \right).
\label{eq:background-subtracted-hamiltonian}
\end{equation}
After applying the barred definitions and Eq.~\eqref{eq:radial-derivative-rescaling}, this becomes
\begin{equation}
 -8\psi^{-5} \left( \psi_{,\bar r\bar r} +\frac{2}{\bar r}\psi_{,\bar r} \right) -6\bar{\mathcal A}^{\,2} = 16\pi \left( \bar\varepsilon_E-\bar\varepsilon_{E\,b}  \right).
\label{eq:barred-subtracted-hamiltonian}
\end{equation}
Thus, neither $\Lambda$ nor the CMC contribution has been omitted from the numerical equations: both cancel against the homogeneous Space II-A background relation. By introducing $u(\bar r) \equiv \psi_{,\bar r} = R_*\psi_{,r}$, the Hamiltonian constraint becomes the first-order pair
\begin{align}
 u(\bar r) &= \psi_{,\bar r} , \label{eq:solver-psi-ode} \\
 u_{,\bar r} &= -\frac{2u}{\bar r} -\frac{\psi^5}{8} \left[ 16\pi \left( \bar\varepsilon_E-\bar\varepsilon_{E\,b} \right) +6\bar{\mathcal A}^{\,2} \right].
\label{eq:solver-hamiltonian-ode}
\end{align}

We next reduce the momentum constraint in Eq.~\eqref{eq:mom}. Because $K$ is spatially constant, this is equivalently
\begin{equation}
 D_j A^{ij} = \frac{8\pi G_0}{c_0^4}j_E^i.
\end{equation}
By using Eq.~\eqref{eq:spherical-tracefree-curvature} and the conformal spatial metric, the covariant divergence evaluates to
\begin{equation}
 D_j A^j{}_r = 2\mathcal A_{,r} +12\frac{\psi_{,r}}{\psi}\mathcal A +\frac{6}{r}\mathcal A. \label{eq:radial-divergence-tracefree}
\end{equation}
Thus, the physical radial momentum equation becomes
\begin{equation}
 2\mathcal A_{,r} +12\frac{\psi_{,r}}{\psi}\mathcal A +\frac{6}{r}\mathcal A = \frac{8\pi G_0}{c_0^4}j_{E\,r}. \label{eq:physical-radial-momentum}
\end{equation}
After conversion to barred variables this becomes
\begin{equation}
 2\bar{\mathcal A}_{,\bar r}+12\frac{\psi_{,\bar r}}{\psi}\bar{\mathcal A}+\frac{6}{\bar r}\bar{\mathcal A} = 8\pi\bar j_{E\,r}.
\label{eq:solver-momentum-residual}
\end{equation}
Solving algebraically for the radial derivative gives the form advanced by the numerical integrator,
\begin{equation}
 \bar{\mathcal A}_{,\bar r} = 4\pi\bar j_{E\,r} - \left( 6\frac{\psi_{,\bar r}}{\psi} +\frac{3}{\bar r} \right) \bar{\mathcal A}. \label{eq:solver-momentum-ode}
\end{equation}

The radial orthonormal basis associated with $\gamma_{rr}=\psi^4$ satisfies
\begin{equation}
\mathbf{e}_{\hat r} = \psi^{-2}\partial_r.
\end{equation}
Consequently, the radial covariant and orthonormal momentum-density components are related by
\begin{equation}
 j_{E\,r} = \psi^2 j_E^{\hat r}, \qquad \bar j_{E\,r} = \psi^2\bar j_E^{\hat r}. \label{eq:radial-source-relation}
\end{equation}

Equations~\eqref{eq:solver-hamiltonian-ode} and \eqref{eq:solver-momentum-ode}, together with Eq.~\eqref{eq:solver-psi-ode}, determine the spherical conformal geometry and trace-free extrinsic curvature for the prescribed matter sources.

The cumulative energy-equivalent mass contrast is integrated simultaneously, but it is not an additional Einstein constraint. From the definition in Eq.~\eqref{eq:energy-compensation},
\begin{equation}
 \Delta M_E(r) = \frac{4\pi}{c_0^2} \int_0^r \left( \varepsilon_E(r')-\varepsilon_{E\,b} \right) \psi^6(r')\,r'^2\,dr',
\end{equation}
one obtains
\begin{equation}
 \frac{d\Delta M_E}{dr} = \frac{4\pi}{c_0^2} \left( \varepsilon_E-\varepsilon_{E\,b} \right) \psi^6 r^2.
\label{eq:physical-mass-contrast-ode}
\end{equation}
Using the convention $\overline{\Delta M}_E = \frac{G_0}{c_0^2R_*}\Delta M_E$, the dimensionless equation of Eq.~\eqref{eq:physical-mass-contrast-ode} becomes
\begin{equation}
 \frac{d \overline{\Delta M}_{E}}{d \bar r} = 4\pi \left( \bar\varepsilon_E-\bar\varepsilon_{E\,b}\right) \psi^6\bar r^2. \label{eq:solver-mass-ode}
\end{equation}
Thus, this integrated state keeps track of cumulative background-relative energy compensation on the chosen ADM slice.

\subsection{Origin regularity and endpoint conditions}
\label{Orireg} 

Regularity at the spherical origin constrains the parity and leading radial behavior of the initial-data variables. A regular scalar such as $\psi$ has an even expansion,
\begin{equation}
 \psi(\bar r) = \psi_c+O(\bar r^2),
\end{equation}
and hence
\begin{equation}
 \psi_{,\bar r}(0)=0.
\end{equation}
A regular radial momentum source must vanish at the origin,
\begin{equation}
 \bar j_{E\,r}=O(\bar r),
\end{equation}
while spherical regularity of the trace-free extrinsic curvature gives
\begin{equation}
 \bar{\mathcal A}=O(\bar r^2).
\end{equation}
Finally, the cumulative mass contrast must vanish with the enclosed volume,
\begin{equation}
 \overline{\Delta M}_E=O(\bar r^3).
\end{equation}
Therefore, the central boundary conditions are 
\begin{equation}
 \psi_{,\bar r}(0)=0, \qquad \bar{\mathcal A}(0)=0, \qquad \overline{\Delta M}_E(0)=0, \qquad \bar j_{E\,r}(0)=0, \label{eq:solver-origin-conditions}
\end{equation}
with finite $\psi(0)$ and $\bar\varepsilon_E(0)$.

For the representative fixed-amplitude solution with $A_-=0.995$, the three shooting parameters are the central conformal factor $\psi(0)$, the dimensionless coordinate endpoint $\bar r_{\rm comp}$, and the velocity-profile amplitude. They are adjusted so that
\begin{equation}
 \psi(\bar r_{\rm comp})-1=0, \qquad \bar{\mathcal A}(\bar r_{\rm comp})=0, \qquad \overline{\Delta M}_E(\bar r_{\rm comp})=0.
\label{eq:solver-three-shooting-conditions}
\end{equation}
These three conditions have distinct roles. The first restores the normalized conformal spatial geometry of the Space II-A FLRW exterior, the second removes the trace-free part of the extrinsic curvature as required by homogeneous FLRW data, and the third enforces vanishing cumulative background-relative energy excess at the endpoint.

The matter-source construction independently supplies
\begin{equation}
 \bar\varepsilon_E(\bar r_{\rm comp}) = \bar\varepsilon_{E\,b}, \qquad
 \bar j_{E\,r}(\bar r_{\rm comp})=0, \qquad
 \bar p(\bar r_{\rm comp})=0, \label{eq:source-endpoint-conditions}
\end{equation}
while the CMC prescription in Eq.~\eqref{eq:cmc} fixes
\begin{equation}
 K=K_{\rm FLRW}=-\frac{3H_p}{c_0}. \label{eq:cmc-endpoint-condition}
\end{equation}
These are source and slicing conditions rather than additional shooting equations.

Exact homogeneous FLRW data also have constant conformal factor, so one
expects
\begin{equation}
 \psi_{,\bar r}(\bar r_{\rm comp})\simeq0.
\label{eq:psi-derivative-closure}
\end{equation}
This condition is deliberately not imposed in the shooting procedure and
therefore provides an independent finite-radius closure diagnostic.

The numerical independent variable is the isotropic coordinate $\bar r$, whereas the physical radius used in the main text is the areal radius $R(r)=\psi^2(r)r$. Unlike the nominal core scale at $\bar r=1$, the outer endpoint is constrained by the shooting condition $\psi(\bar r_{\rm comp})=1$. Therefore,
\begin{equation}
 R_{\rm comp}=r_{\rm comp}. \label{eq:coordinate-areal-endpoint}
\end{equation}
Since $R_*=\Rvir$, the dimensionless endpoint reported in the main text is consequently
\begin{equation}
 \bar r_{\rm comp}= \frac{r_{\rm comp}}{R_*}=\frac{\Rcomp}{\Rvir}. \label{eq:dimensionless-compensation-radius}
\end{equation}
 
Therefore, the endpoint construction separates three logically distinct requirements: the conditions solved by shooting, Eq.~\eqref{eq:solver-three-shooting-conditions}; the matter and CMC endpoint conditions supplied by the source and slicing prescriptions, Eqs.~\eqref{eq:source-endpoint-conditions} and \eqref{eq:cmc-endpoint-condition}; and the unused derivative condition in Eq.~\eqref{eq:psi-derivative-closure}, which serves as an independent check of closure to the exact Space II-A FLRW initial data.

\FloatBarrier

\section{Detailed numerical diagnostics and reproducibility}
\label{app:admissibility} 

Appendix~\ref{app:spherical-constraints} derives the spherical conformal ADM initial-data equations and endpoint conditions for the prescribed matter sources. We test the resulting numerical solution in four complementary ways. First, we evaluate the Hamiltonian and momentum constraints on the computed state. Second, we test whether the geometry, extrinsic curvature, and matter sources close together to the homogeneous Space II-A FLRW initial data at a finite radius. Third, we check whether the smooth positive-density compensation also satisfies the local weak, null, and dominant energy-condition inequalities. Fourth, separately from the slice-dependent proper-volume compensation, we reconstruct the invariant Misner--Sharp mass length and test its endpoint closure. These diagnostics address different admissibility properties and are therefore not interchangeable. The final subsection specifies a reproducible numerical protocol for all four tests.

All quantities below use the barred normalization of Appendix~\ref{app:conventions-normalization}; barred residuals are dimensionless rather than SI values. The scope throughout is one local-effective initial slice with $c_0$ and $G_0$ frozen. No time evolution, formation or virialization mechanism, stability or persistence result, radial interpolation of $c$ or $G$, inhomogeneous GCT lapse, or full global-GCT junction/evolution is constructed here.

\subsection{Hamiltonian and momentum constraints}
\label{app:hamilandmom} 

Positive density alone does not define admissible relativistic initial data. The spatial geometry and extrinsic curvature must satisfy the Hamiltonian and momentum constraints. Thus, we define the physical constraint residuals by
\begin{align}
\mathcal H &\equiv {}^{(3)}R+K^2-K_{ij}K^{ij} -\frac{16\pi G_0}{c_0^4}\varepsilon_E -2\Lambda , \label{eq:Hresdef} \\
\mathcal M^i &\equiv D_j\!\left(K^{ij}-\gamma^{ij}K\right) -\frac{8\pi G_0}{c_0^4}j_E^i . \label{eq:Mresdef}
\end{align}
The Hamiltonian residual $\mathcal H$ tests the balance between the spatial geometry and the Eulerian energy source, whereas $\mathcal M^i$ measures the corresponding momentum-constraint residual. In spherical symmetry, we evaluate its covariant radial component $\mathcal M_r$, which tests the trace-free-curvature balance against the radial momentum source.

\begin{figure}[!htbp]
\centering
\includegraphics[width=0.72\linewidth]{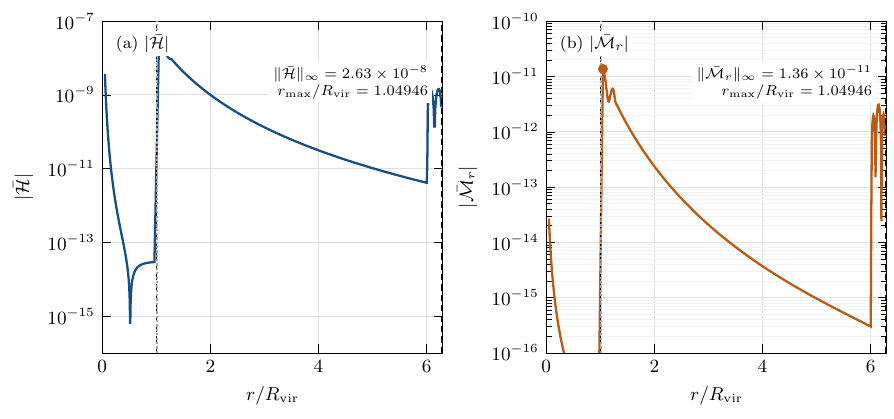}
\caption{Independent barred Hamiltonian and covariant radial momentum constraint residuals for the representative $N=800$ solution. The left and right panels show $|\bar{\mathcal H}|$ and $|\bar{\mathcal M}_{r}|$, respectively. Vertical markers denote the nominal virial/core scale $\bar r=1$, the shoulder endpoint $\bar r=1.01$, and the finite-radius FLRW closure endpoint $R_{\rm comp}$. Filled circles mark the bulk maxima.}
\label{fig:constraints}
\end{figure}

Using the barred variables of Appendix~\ref{app:conventions-normalization} and the background subtraction derived in Appendix~\ref{app:spherical-constraints}, the dimensionless Hamiltonian residual evaluated for Fig.~\ref{fig:constraints} is
\begin{equation}
\bar{\mathcal H} \equiv R_*^2\mathcal H = -8\psi^{-5} \left( \psi_{,\bar r\bar r} +\frac{2}{\bar r}\psi_{,\bar r} \right) -6\bar{\mathcal A}^{2}
-16\pi \left( \bar\varepsilon_E-\bar\varepsilon_{E\,b} \right). \label{eq:Hresbarred}
\end{equation}
The corresponding dimensionless covariant radial momentum residual is
\begin{equation}
\bar{\mathcal M}_{r} \equiv R_*^2\mathcal M_r = 2\bar{\mathcal A}_{,\bar r} +12\frac{\psi_{,\bar r}}{\psi}\bar{\mathcal A} +\frac{6}{\bar r}\bar{\mathcal A} -8\pi\bar j_{E\,r},
\label{eq:Mresbarred}
\end{equation}
where the subscript $r$ labels the covariant radial component, whereas a comma followed by $r$ or $\bar r$ denotes differentiation.

The resolution label $N$ denotes the total number of radial integration intervals used to solve the numerical initial-data problem. Thus, for a single ordered radial grid $\{\bar r_i\}_{i=0}^{N}$, there are $N$ intervals $[\bar r_i,\bar r_{i+1}]$ and $N+1$ grid points. The numerical domain is partitioned into piecewise radial subdomains so that the prescribed source-transition radii are explicit subdomain boundaries. Refining $N$ therefore increases the radial resolution without allowing an integration step to smear a prescribed transition across a numerical cell.

For each resolution, the first-order system derived in Appendix~\ref{app:spherical-constraints},
\begin{equation}
\left\{ \psi,\, u\equiv\psi_{,\bar r},\, \bar{\mathcal A},\, \overline{\Delta M}_E \right\},
\end{equation}
is integrated radially outward from the regular center. At a trial set of shooting parameters, this integration produces endpoint values for the three residual conditions in Eq.~\eqref{eq:solver-three-shooting-conditions}. The central conformal factor, the compensation endpoint, and the velocity-profile amplitude are then adjusted, and the radial integration is repeated until
\begin{equation}
\psi(\bar r_{\rm comp})-1=0, \qquad \bar{\mathcal A}(\bar r_{\rm comp})=0, \qquad \overline{\Delta M}_E(\bar r_{\rm comp})=0
\end{equation}
are simultaneously satisfied to the prescribed numerical tolerances. Thus, $N=800$ refers to a fully solved coupled radial initial-data and shooting problem, not to the sampling of a pre-existing analytic profile. The shooting problem itself is solved again independently for $N=200,400,800,$ and $1600$; the $N=800$ solution is used for the representative profiles, while the full sequence tests resolution convergence.

The first-order system is advanced with fourth-order Runge--Kutta integration. Once a converged shooting solution has been obtained, however, Eqs.~\eqref{eq:Hresbarred} and \eqref{eq:Mresbarred} are evaluated \emph{a posteriori} using derivatives reconstructed from the stored neighboring radial values with a separate centered finite-difference operator. No second radial integration is performed at this stage: the residuals are reconstructed algebraically from the already obtained numerical profiles and independently evaluated radial derivatives. They are therefore not simply the right-hand sides used by the RK4 integrator printed back as a diagnostic.

A centered finite-difference stencil, i.e.\ the neighboring grid points used to reconstruct a derivative, is not an appropriate bulk estimator at the coordinate origin or when it crosses a prescribed piecewise source-transition boundary. Such stencils are excluded by a resolution-independent, predeclared bulk-mask rule. The reported $L^\infty$ and $L^2$ norms are therefore evaluated on a fixed class of regular bulk points, rather than after removing grid points according to the size of their residuals. Since this independent residual evaluator is second order, second-order convergence of the residuals is consistent with the use of a fourth-order RK4 integrator for the underlying initial-data equations.

Figure~\ref{fig:constraints} shows the radial profiles obtained from this independent residual evaluation. Both constraints remain small over the numerical domain. For the representative $N=800$ solution, the largest declared bulk residuals are
\begin{equation}
\|\bar{\mathcal H}\|_\infty = 2.63\times10^{-8}, \qquad
\|\bar{\mathcal M}_{r}\|_\infty = 1.36\times10^{-11}.
\end{equation}
Both maxima occur at
\begin{equation}
\bar r_{\rm max}\simeq1.04946,
\end{equation}
just outside the narrow shoulder. This is the part of the source profile with the most rapid radial variation, and the independently reconstructed finite-difference derivatives are correspondingly most sensitive there. The localized maxima therefore identify the most demanding region for the discrete residual evaluator; they do not indicate a physical singularity or a breakdown of the constraints. Over most of the broader compensation region the residuals are substantially smaller.

The more important consistency test is the behavior of these residuals under radial refinement. As shown in Table~\ref{tab:constraints}, the complete shooting problem is solved independently at $N=200,400,800,$ and $1600$ before the residual evaluator is applied. For the Hamiltonian residual, the coarse $200\to400$ reduction gives an apparent convergence order of about $3.8$. This coarse pair is pre-asymptotic and is therefore not interpreted as a continuum convergence rate. The subsequent effective orders are approximately $2.2$ for $400\to800$ and $2.0$ for $800\to1600$, approaching the second-order behavior expected from the centered finite-difference residual evaluator. The momentum residual shows the same approximately second-order refinement behavior over the resolved sequence. The refinement study therefore establishes numerical constraint satisfaction of the initial-data solution to the reported accuracy; it is not a test of dynamical formation or stability.

\FloatBarrier

\subsection{Finite-radius FLRW matching}
\label{app:finradFLRW} 

Density matching alone is not sufficient for finite-radius closure of the initial data. The conformal geometry, extrinsic curvature, and matter sources must approach together the values induced by the homogeneous Space II-A FLRW exterior. The corresponding endpoint values are
\begin{equation}
\psi=1,\qquad \psi_{,\bar r}=0,\qquad \bar{\mathcal A}=0,\qquad \bar j_{E\,r}=0,\qquad \varepsilon_E=\varepsilon_{E\,b},\qquad p=0,\qquad
K=K_{\rm FLRW} \qquad \text{at } R=R_{\rm comp}, \label{eq:allmatching}
\end{equation}
where $R=\psi^2r$ is the areal radius. Because the shooting condition imposes $\psi=1$ at the outer endpoint, the isotropic and areal radii coincide there, $R_{\rm comp}=r_{\rm comp}$, as shown in Eq.~\eqref{eq:coordinate-areal-endpoint}. The value $\psi_{,\bar r}=0$ in Eq.~\eqref{eq:allmatching} is the homogeneous FLRW target value; it is not imposed as a shooting condition and thus provides an independent geometric closure diagnostic.

\begin{figure}[!htbp]
\centering
\includegraphics[width=0.72\linewidth]{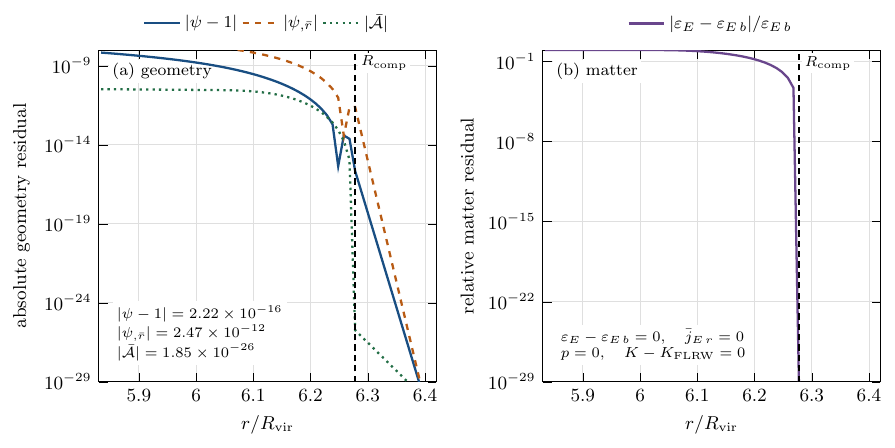}
\caption{Finite-radius return to the Space II-A FLRW initial data, shown over the outer compensation sector. The left panel displays $|\psi-1|$, $|\psi_{,\bar r}|$, and $|\bar{\mathcal A}|$, while the right panel shows the relative Eulerian-energy mismatch in Eq.~\eqref{eq:relative-energy-matching}. The vertical marker denotes the finite-radius closure endpoint $R_{\rm comp}$.}
\label{fig:matching}
\end{figure}

Figure~\ref{fig:matching} separates the barred geometric diagnostics
\begin{equation}
|\psi-1|,\qquad |\psi_{,\bar r}|,\qquad |\bar{\mathcal A}| \label{eq:geometry-matching-diagnostics}
\end{equation}
from the relative Eulerian-energy mismatch
\begin{equation}
\frac{|\bar\varepsilon_E-\bar\varepsilon_{E\,b}|}{\bar\varepsilon_{E\,b}} =
\frac{|\varepsilon_E-\varepsilon_{E\,b}|}{\varepsilon_{E\,b}}. \label{eq:relative-energy-matching}
\end{equation}

Figure~\ref{fig:matching} shows two complementary aspects of the finite-radius closure. In the left panel, the conformal-factor residual, the independently tested radial derivative, and the trace-free extrinsic-curvature amplitude all approach their homogeneous FLRW values as the endpoint is reached. For the representative solution,
\begin{equation}
|\psi-1| = 2.22\times10^{-16}, \qquad |\psi_{,\bar r}| = 2.47\times10^{-12}, \qquad
|\bar{\mathcal A}| = 1.85\times10^{-26} \qquad  \text{at } R=R_{\rm comp}. \label{eq:matching-geometry-endpoint}
\end{equation}
At the same endpoint, the source construction gives
\begin{equation}
\bar j_{E\,r}=0,\qquad \varepsilon_E=\varepsilon_{E\,b},\qquad p=0, \qquad K=K_{\rm FLRW}. \label{eq:matching-source-endpoint}
\end{equation}
Thus, the conformal geometry, extrinsic curvature, and matter sources return together to the Space II-A exterior data rather than only the density being matched.

The right panel displays a different aspect of the construction. The relative Eulerian-energy mismatch is not expected to vanish throughout the displayed interval because the compensation sector is deliberately inhomogeneous. It is approximately $0.995$ at $\bar r\simeq5.834$, decreases to $0.954$ at $\bar r\simeq6.061$, and reaches $4.94\times10^{-4}$ at $\bar r\simeq6.268$, immediately before $R_{\rm comp}/R_{\rm vir} = 6.27789$. At the endpoint, the prescribed condition $\varepsilon_E=\varepsilon_{E\,b}$ makes the relative mismatch reach the numerical floor. Therefore, the figure does not represent a homogeneous exterior throughout the region inside $R_{\rm comp}$. Rather, it shows a finite-width inhomogeneous compensation sector that returns to homogeneous Space II-A initial data at a finite radius.

Taken together, the geometric and matter endpoint checks show that the numerically constructed Space II-B portion closes to the analytic Space II-A initial data without a residual distributional layer on the constructed spacelike hypersurface. This statement concerns the initial slice only and is not a claim about an evolved timelike junction.

\FloatBarrier

\subsection{Local energy-condition audit}
\label{app:locenecon} 

The radial density morphology is already shown in Fig.~\ref{fig:density}. The separate question here is whether the same isotropic perfect-fluid rest-frame source satisfies the tested pointwise energy conditions. At every stored radial point we evaluate
\begin{equation}
q_e\equiv\frac{e}{e_b},\qquad q_{\rm N}\equiv\frac{e+p}{e_b},\qquad q_{\rm D}\equiv\frac{e-|p|}{e_b}. \label{eq:ECmargins}
\end{equation}
Because $e_b>0$, WEC requires $q_e,q_{\rm N}\ge0$, NEC requires $q_{\rm N}\ge0$, and DEC requires $q_{\rm D}\ge0$. Table~\ref{tab:energy-condition-audit} reports the minima over the numerical domain.

\begin{table}[!htbp]
\centering
\caption{Pointwise energy-condition audit for the representative $N=800$ solution. The pass threshold for every listed normalized margin is zero.}
\begin{tabular}{lcc}
\toprule
Margin & Minimum & Location $\bar r$\\
\midrule
$e/e_b$ & $0.00499994$ & $6.01154$\\
$(e+p)/e_b$ & $0.00499994$ & $6.01154$\\
$(e-|p|)/e_b$ & $0.00499994$ & $6.01154$\\
\bottomrule
\end{tabular}
\label{tab:energy-condition-audit}
\end{table}

All three normalized margins remain strictly positive. Thus, the smooth compensation does not require exotic bulk matter under the adopted isotropic perfect-fluid closure. This is an initial-slice matter-admissibility test: the strong energy condition is not tested, and no formation, stability, or persistence conclusion follows.

\FloatBarrier

\subsection{Invariant Misner--Sharp mass-length diagnostic}
\label{app:invmisshr} 

Proper-volume compensation is slice dependent. In particular, $\Delta M_E$ integrates the Eulerian energy contrast with the proper-volume element of the chosen ADM slice. A geometrically distinct spherical check is provided by the Misner--Sharp mass length $\mu_{\rm MS}$, defined with the vacuum-$\Lambda$ contribution kept explicit as
\begin{equation}
1-\frac{2\mu_{\rm MS}}{R}-\frac{\Lambda R^2}{3} = g^{ab}\partial_aR\,\partial_bR, \qquad R=\psi^2r. \label{eq:msappendix}
\end{equation}
Because $\mu_{\rm MS}$ is reconstructed from the spherical spacetime geometry, it is invariant under spherical-coordinate reparametrizations. This diagnostic is distinct from the proper-volume quantity $\Delta M_E$, and the two need not agree pointwise at intermediate radii: pressure, binding, geometric, and slicing information enter them differently. Their endpoint closure tests must therefore also remain distinct.

\begin{figure}[!htbp]
\centering
\includegraphics[width=0.90\linewidth,height=0.55\textheight,keepaspectratio]{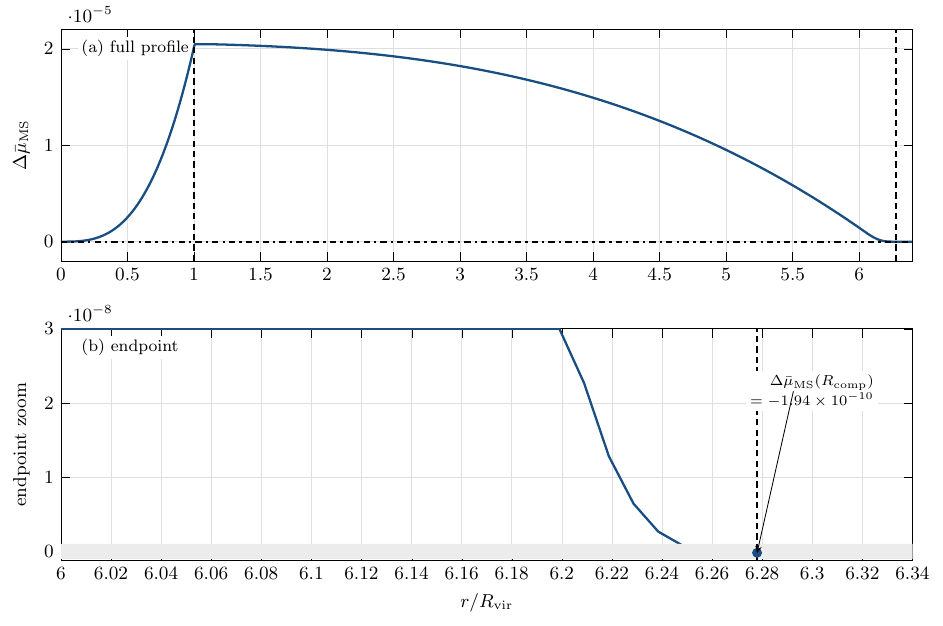}
\caption{Background-relative invariant Misner--Sharp mass-length contrast $\Delta\bar\mu_{\rm MS}$ for the representative solution. The upper panel shows the full radial profile, while the lower panel magnifies the finite-radius closure region. Vertical markers denote the nominal core scale $\bar r=1$ and $R_{\rm comp}$; the gray band in the lower panel indicates the declared endpoint tolerance $|\Delta\bar\mu_{\rm MS}|\le10^{-9}$.}
\label{fig:misnersharp}
\end{figure}

Figure~\ref{fig:misnersharp} shows the background-relative Misner--Sharp mass-length contrast reconstructed independently from the spherical geometry. The nonzero profile at intermediate radii reflects the inhomogeneous geometry of the core and compensation region and is not a failure of compensation. In particular, there is no requirement that $\Delta\bar\mu_{\rm MS}$ follow the slice-dependent proper-volume quantity $\Delta M_E$ point by point.

The lower panel tests the quantity relevant for finite-radius invariant closure. At the numerical endpoint,
\[\Delta\bar\mu_{\rm MS}(R_{\rm comp}) \simeq -1.94\times10^{-10}, \]
which lies within the declared tolerance $|\Delta\bar\mu_{\rm MS}|\le10^{-9}$. The small remaining endpoint offset is sensitive to the nonzero geometric residual $\psi_{,\bar r}(\bar r_{\rm comp})$. Since $R=\psi^2r$,
\[R_{,r}=\psi^2+2\psi r\psi_{,r},\]
and this derivative enters the invariant combination in Eq.~\eqref{eq:msappendix}. Therefore, the residual $\Delta\bar\mu_{\rm MS}(R_{\rm comp})$ does not indicate a failure of the separately imposed proper-volume condition $\Delta M_E(r_{\rm comp})=0$. Conversely, that imposed condition alone would not establish invariant mass-length closure; the lower panel of Fig.~\ref{fig:misnersharp} provides this additional geometric check.

For physical interpretation, $\mu_{\rm MS}$ is an invariant mass length rather than a mass in kilograms. As summarized in Appendix~\ref{app:conventions-normalization}, the corresponding kilogram conversion depends on the stated sector:
\[
m_{\rm MS}^{\rm loc} = \frac{c_0^2}{G_0}\mu_{\rm MS}, \qquad
m_{\rm MS}^{\rm bg}(T) = \frac{c_{\rm bg}^2(T)}{G_{\rm bg}(T)}\mu_{\rm MS}.
\]
Thus, the invariant length is the common geometric diagnostic, whereas a mass in kilograms requires the specified local-frozen or global-background conversion convention.

An independent manufactured-state mass test validates the dimensionless geometric combination $g^{ab}\partial_aR\,\partial_bR$ entering Eq.~\eqref{eq:msappendix}, and hence the numerical reconstruction of $\mu_{\rm MS}$. This is a spatial/geometric reconstruction test. It neither selects a kilogram-conversion convention nor tests physical time evolution.

\FloatBarrier

\subsection{Numerical procedure and reproducibility protocol}
\label{app:reproducibility} 

The calculation begins by applying the barred normalization of Appendix~\ref{app:conventions-normalization}, with $R_*=\Rvir$, to a spherical radial domain extending from the regular center to the unknown finite FLRW closure radius. A prescribed positive compensated matter-source family supplies the Eulerian energy and momentum sources together with the rest-frame energy density and pressure. The domain is partitioned so that each prescribed source-transition radius is an explicit grid boundary. This prevents an integration step from crossing a prescribed source-transition boundary and preserves the same piecewise source structure at every resolution.

At fixed $A_-=0.995$, the first-order Hamiltonian, momentum, geometry, and cumulative-compensation system derived in Appendix~\ref{app:spherical-constraints} is integrated radially outward with a fourth-order Runge--Kutta method. Regular origin expansions initialize $\psi$, $u=\psi_{,\bar r}$, $\bar{\mathcal A}$, and $\overline{\Delta M}_E$. Simultaneous root finding determines the central conformal factor, compensation endpoint, and velocity-profile amplitude so that
\[
\psi(\bar r_{\rm comp})=1,\qquad \bar{\mathcal A}(\bar r_{\rm comp})=0,\qquad \overline{\Delta M}_E(\bar r_{\rm comp})=0 .
\]
The resulting $N=800$ solution is stored as the representative numerical state. For the resolution study, the complete shooting problem---not merely a resampling of the final profile---is solved independently at $N=200,400,800,$ and $1600$ radial intervals.

Verification is performed after each shooting solution has been obtained. The Hamiltonian and covariant radial momentum residuals are reconstructed with the separate second-order centered finite-difference evaluator and the predeclared bulk mask defined in Sec.~\ref{app:hamilandmom}. Endpoint closure is then checked jointly for the conformal geometry, trace-free and CMC extrinsic curvature, Eulerian energy density, covariant radial momentum density, pressure, and cumulative proper-volume compensation, as described in Sec.~\ref{app:finradFLRW}. The local rest-frame stress tensor is independently evaluated for the WEC, NEC, and DEC margins of Sec.~\ref{app:locenecon}, while the Misner--Sharp mass length is reconstructed from the spherical geometry as described in Sec.~\ref{app:invmisshr}. Thus, no single diagnostic is used as a proxy for another.

The barred-to-SI restoration map of Eq.~\eqref{eq:bars} is checked in both directions, including the distinction between the local frozen-constant convention and the global-background mass-conversion convention, to a relative tolerance of $10^{-12}$. Manufactured-state, null-background, and operator tests independently check the normalization, spherical spatial operators, origin treatment, constraint evaluator, and invariant-mass reconstruction. For example, the independent manufactured-state $400\to800$ orders are
\[
q_R=3.9376,\qquad q_{\mathcal H}=3.9376,\qquad q_{\mathcal M_r}=3.9998,\qquad q_{R,\rm origin}=4.4837 .
\]
These manufactured-state orders characterize the dedicated operator tests and should not be identified with the second-order \emph{a posteriori} constraint-residual convergence of the numerical shooting solutions discussed in Sec.~\ref{app:hamilandmom}. They test the numerical spatial operators and normalization rather than physical time evolution.

The plotted curves are generated from the stored representative numerical profiles only after these checks have been completed; figure preparation does not solve the field equations again. Root tolerances, residual normalizations, full-precision outputs, and the provenance of the rounded publication values are retained in the machine-readable archive. For exact identification of the stored $N=800$ state, the archived records give
\[
\frac{\Rcomp}{\Rvir} = 6.277892900604803, \qquad \min\!\left(\frac{e}{e_b}\right) = 0.004999936409861388,
\]
\[
\|\bar{\mathcal H}\|_\infty = 2.6316378805277608\times10^{-8}, \qquad \|\bar{\mathcal M}_r\|_\infty = 1.3645518983153954\times10^{-11},
\]
\[
\psi_{,\bar r}(\bar r_{\rm comp}) =2.4671659737550144\times10^{-12}, \qquad \Delta\bar\mu_{\rm MS}(R_{\rm comp}) =-1.9447196890020582\times10^{-10}.
\]
These digits identify a particular archived numerical state and should not be interpreted as continuum-level physical accuracy.

The present protocol therefore reconstructs and tests one constraint-satisfying initial slice only; it contains no time integration. Dynamical collapse, the formation of a virialized core, and the subsequent persistence or stability of the constructed local sector constitute a separate evolution problem and lie outside the scope of the present initial-data calculation.

\FloatBarrier


\end{document}